\documentclass[12pt]{article}

\usepackage{amsmath}
\usepackage{amssymb}
\usepackage[backend=bibtex,doi=false,sorting=none,firstinits=true]{biblatex}
\usepackage[T1]{fontenc}
\usepackage[utf8]{inputenc}
\usepackage{lmodern}
\usepackage{mathtools}
\usepackage{slashed}
\usepackage{subfig}
\usepackage{titlesec}

\usepackage{hyperref}

\if@twoside \oddsidemargin -17pt \evensidemargin 00pt
\else \oddsidemargin 00pt \evensidemargin 00pt
\fi
\renewbibmacro{in:}{}

\AtEveryBibitem{\clearfield{title}}

\makeatletter

\newcommand\RedeclareMathOperator{%
\@ifstar{\def\rmo@s{m}\rmo@redeclare}{\def\rmo@s{o}\rmo@redeclare}%
}

\newcommand\rmo@redeclare[2]{%
\begingroup\escapechar\m@ne\xdef\@gtempa{{\string#1}}\endgroup
\expandafter\@ifundefined\@gtempa{\@latex@error{\noexpand#1undefined}\@ehc}%
\relax
\expandafter\rmo@declmathop\rmo@s{#1}{#2}}
\newcommand\rmo@declmathop[3]{\DeclareRobustCommand{#2}{\qopname\newmcodes@#1{#3}}}
\@onlypreamble\RedeclareMathOperator

\def\dsl{\mathpalette\make@slash}
\def\make@slash#1#2{\setbox\z@\hbox{$#1#2$}%
  \hbox to 0pt{\hss$#1/$\hss\kern-\wd0}\box0}

\makeatother

\renewenvironment{abstract}{{\indent\bfseries\abstractname:}\par}

\RedeclareMathOperator{\Re}{Re}

\DeclareMathOperator{\Li2}{Li_2}
\DeclareMathOperator{\cLi2}{{\cal L}i_2}

\titleformat*{\subsubsection}{\large\it}

\numberwithin{equation}{section}

\renewcommand{\vec}[1]{\boldsymbol{#1}}

\def\citere#1{\mbox{Ref.~\cite{#1}}}

\def\mathswitch#1{\relax\ifmmode#1\else$#1$\fi}
\def\mathswitchr#1{\relax\ifmmode{\mathrm{#1}}\else$\mathrm{#1}$\fi}
\def\mathswitchit#1{\relax\ifmmode{#1}\else$#1$\fi}

\newcommand{\PW}{\mathswitchr W}
\newcommand{\PZ}{\mathswitchr Z}

\newcommand{\Pd}{\mathswitchr d}

\newcommand{\Pu}{\mathswitchr u}

\newcommand{\Pp}{\mathswitchr p}

\newcommand{\Pq}{\mathswitchit q}
\newcommand{\Pep}{\mathswitchr {e^+}}
\newcommand{\Pem}{\mathswitchr {e^-}}

\newcommand{\PWp}{\mathswitchr {W^+}}
\newcommand{\PWm}{\mathswitchr {W^-}}
\newcommand{\PWpm}{\mathswitchr {W^\pm}}

\newcommand{\MW}{\mathswitch {M_\PW}}

\newcommand{\pro}{{\mathrm{prod}}}
\newcommand{\dec}{{\mathrm{dec}}}
\newcommand{\out}{{\mathrm{out}}}
\newcommand{\inc}{{\mathrm{in}}}
\newcommand{\eik}{{\mathrm{eik}}}

\newcommand{\ri}{{\mathrm{i}}}

\newcommand{\ieps}{\ri\epsilon}
\newcommand{\rd}{{\mathrm{d}}}

\def\draftdate{\relax}
\def\mda{\relax}
\def\mua{\relax}
\def\mla{\relax}
\def\draft{
\def\thtystars{******************************}
\def\sixtystars{\thtystars\thtystars}
\typeout{}
\typeout{\sixtystars**}
\typeout{* Draft mode!
         For final version remove \protect\draft\space in source file *}
\typeout{\sixtystars**}
\typeout{}
\def\draftdate{\today}
\def\mua{\marginpar[\boldmath\hfil$\uparrow$]%
                   {\boldmath$\uparrow$\hfil}%
                    \typeout{marginpar: $\uparrow$}\ignorespaces}
\def\mda{\marginpar[\boldmath\hfil$\downarrow$]%
                   {\boldmath$\downarrow$\hfil}%
                    \typeout{marginpar: $\downarrow$}\ignorespaces}
\def\mla{\marginpar[\boldmath\hfil$\rightarrow$]%
                   {\boldmath$\leftarrow $\hfil}%
                    \typeout{marginpar: $\leftrightarrow$}\ignorespaces}
\def\Mua{\marginpar[\boldmath\hfil$\Uparrow$]%
                   {\boldmath$\Uparrow$\hfil}%
                    \typeout{marginpar: $\Uparrow$}\ignorespaces}
\def\Mda{\marginpar[\boldmath\hfil$\Downarrow$]%
                   {\boldmath$\Downarrow$\hfil}%
                    \typeout{marginpar: $\Downarrow$}\ignorespaces}
\def\Mla{\marginpar[\boldmath\hfil$\Rightarrow$]%
                   {\boldmath$\Leftarrow $\hfil}%
                    \typeout{marginpar: $\Leftrightarrow$}\ignorespaces}
\overfullrule 5pt
\oddsidemargin -15mm
\marginparwidth 29mm
}

\begin{document}

\thispagestyle{empty}

\hfill FR-PHENO-2015-008
\vfill
\begin{center}
  {\Large \textbf{Non-factorizable photonic corrections to resonant production
                  and decay of many unstable particles}
    \par} \vskip 2.5em
  {\large
    {\sc Stefan Dittmaier and Christopher Schwan
    }\\[1ex]
    {\normalsize \textit{Albert-Ludwigs-Universit\"at Freiburg, Physikalisches
                         Institut, \\ D-79104 Freiburg, Germany}
    }
    \\[2ex]
  }
  \par \vskip 1em
\end{center}
\vfill

\begin{abstract}
Electroweak radiative corrections to the production of high-multiplicity final
states with several intermediate resonances in most cases can be sufficiently
well described by the leading contribution of an expansion about the resonance
poles. In this approach, also known as pole approximation, corrections are
classified into separately gauge-invariant factorizable and non-factorizable
corrections, where the former can be attributed to the production and decay of
the unstable particles on their mass shell. The remaining non-factorizable
corrections are induced by the exchange of soft photons between different
production and decay subprocesses. We give explicit analytical results for the
non-factorizable photonic virtual corrections to the production of an arbitrary
number of unstable particles at the one-loop level and, thus, deliver an
essential building block in the calculation of next-to-leading-order electroweak
corrections in pole approximation. The remaining virtual factorizable
corrections can be obtained with modern automated one-loop matrix-element
generators, while the evaluation of the corresponding real photonic corrections
can be evaluated with full matrix elements by multi-purpose Monte Carlo
generators. Our results can be easily modified to non-factorizable QCD
corrections, which are induced by soft-gluon exchange.
\end{abstract}

\vskip 2cm
\noindent\textbf{November 2015}

\setcounter{page}{0}
\clearpage

\section{Introduction}

With very few exceptions, all interesting fundamental particles are unstable
and can only be reconstructed after collecting their decay products in
detectors.
In the Standard Model (SM), this most notably concerns the gauge bosons
$\PW$ and $\PZ$ of the weak interaction, the top quark, and the Higgs boson,
for which a candidate was found at the LHC in 2012.
In extensions of the SM, typically more heavy, unstable particles are predicted,
such as additional Higgs bosons or
gluinos, charginos, neutralinos, and sfermions in supersymmetric theories.
After the first period of data taking at the LHC, the SM is in better shape than
ever in describing practically all phenomena in high-energy particle physics.
The search for new physics, thus, has to proceed with precision at the highest possible
level, in order to reveal any possible deviation from SM predictions.
To this end, both QCD and electroweak corrections have to be included in cross-section predictions.

Production processes of unstable particles notoriously lead to many-particle final
states where the bulk of cross-section contributions results from phase-space
regions where the intermediate unstable particles are resonant, i.e.\ near their
mass shell. In the usual perturbative evaluation of scattering amplitudes in
quantum field theory, a particle propagator develops a pole at the resonance point,
i.e.\ a proper resonance description requires at least a partial resummation of
self-energy corrections to the propagator near the resonance.
Since this procedure mixes perturbative orders, such Dyson
summations potentially
lead to violations of identities (Ward, Slavnov--Taylor, Nielsen identities)
that manifest gauge invariance order by order in perturbation theory.
A more detailed discussion of this issue and further references can be found
in Ref.~\cite{Butterworth:2014efa}.
The two most prominent procedures to avoid the gauge breaking are
the so-called {\it pole scheme}~\cite{Stuart:1991xk,Aeppli:1993rs}
and the {\it complex-mass scheme}~\cite{Denner:1999gp,Denner:2005fg}.
Both make use of the fact that the complex pole location $p^2=\overline{M}^2$
of an unstable particle's propagator with momentum transfer $p$
is a gauge-invariant quantity which
can serve for a proper mass and decay width definition~\cite{Gambino:1999ai,Grassi:2001bz}.
In the complex-mass scheme the complex masses are consistently introduced
as input parameters, so that all coupling parameters derived from the masses,
like the electroweak mixing angle in the SM, become complex. Being a consistent
analytical continuation to complex parameters, this scheme fully maintains
gauge invariance.%
\footnote{The complex-mass scheme introduces spurious unitarity violation,
which is, however, always beyond the level of completely calculated
orders~\cite{Denner:2014zga}, i.e.\ the spurious terms are
of next-to-next-to-leading order in next-to-leading order calculations, etc.}
The scheme delivers the same level of accuracy in resonant and non-resonant regions
in phase space. However, if one is only interested in the resonance regions,
which is typically the case in many-particle processes with low cross sections,
the scheme leads to a proliferation of terms induced by the numerous Feynman diagrams
contributing only in off-shell regions.

The pole scheme suggests to isolate the gauge-invariant residues of the resonance
poles and to introduce propagators with complex masses $\overline{M}$ only there, while keeping
the remaining parts untouched. Restricting this general procedure to resonant contributions
defines the {\it pole approximation} (PA), which is adequate if only the off-shell
behaviour of cross sections near resonances is relevant, but contributions deep in the
off-shell region are negligible.
The corrections to the resonance residues comprise the corrections to the
production and decay subprocesses with on-shell kinematics for the resonant particles.
Since these contributions to matrix elements contain explicit resonance factors
$\propto 1/(p^2-\overline{M}^2)$, they are called {\it factorizable} corrections.
The remaining resonant contribution in the PA furnish the
{\it non-factorizable} corrections.
They result from the fact that the infrared (IR) limit in loop diagrams
and in real emission contributions and the procedure of setting particle momenta
on their mass shell do not commute with each other if the on-shell limit leads
to soft IR singularities. This is the case if a soft (real or virtual) massless gauge boson
bridges a resonance. The fact that only the soft momentum region of the massless gauge
boson leads to resonant contribution simplifies the calculation of the
non-factorizable corrections, because factorization properties of the underlying
diagrams can be exploited. The terminology ``non-factorizable'', thus, does not refer to
factorization properties of diagrammatic parts, but to
the off-shell behaviour of the corrections, which apart from resonance
factors $1/(p^2-\overline{M}^2)$ contain non-analytic terms like $\ln(p^2-\overline{M}^2)$.

The complex-mass and pole schemes were successfully used in many higher-order calculations,
both for electroweak and QCD corrections. Here we just mention the two examples of
single and pair production of the weak gauge bosons $\PW$ and $\PZ$, where
results of the two schemes have been compared in detail.
For W-pair production at LEP2, $\Pep\Pem\to\PW\PW\to4\,$fermions,
the {\it double-pole approximation} (DPA) for the two $\PW$~resonances was worked out
in different next-to-leading-order (NLO)
variants~\cite{Beenakker:1998gr,Jadach:1998tz,Denner:1999kn,Denner:2000bj},
which were numerically compared in detail~\cite{Grunewald:2000ju}.
Later the comparison to the full off-shell NLO
calculation~\cite{Denner:2005fg,Denner:2005es}
within the complex-mass scheme confirmed both the expected accuracy of the DPA in the
resonance region and the limitation in the transition region to the off-shell domains.
The situation is expected to be similar for W-pair production at hadron colliders,
where up to now only results in DPA are known~\cite{Accomando:2004de,Billoni:2013aba}.
For the conceptionally simpler Drell--Yan process of single $\PW/\PZ$ production at
hadron colliders, detailed comparisons between PA and
complex-mass scheme are discussed in Refs.~\cite{Dittmaier:2001ay,Dittmaier:2014qza}.
In Ref.~\cite{Dittmaier:2014qza}
the concept of a PA was carried to the next-to-next-to-leading-order 
level and applied to
the mixed QCD--electroweak corrections of ${\cal O}(\alpha_{\mathrm{s}}\alpha)$.
Applications of the PA to processes with more than two
resonances only exist for leading-order (LO) predictions
(see, e.g., \citere{Denner:2014wka}).

The concept of the PA can be carried out both for virtual and
real radiative corrections, however, care has to be taken that the approximations
are set up in such a way that the cancellation of IR (soft and/or collinear)
singularities between virtual and real corrections is not disturbed.
If both virtual and real corrections are treated in PA, the
sum of virtual and real non-factorizable corrections forms a closed gauge-invariant,
IR-finite subset of corrections that can be discussed separately.
For single and double resonances
it has been shown that these completely cancel at NLO~\cite{Fadin:1993dz,Melnikov:1995fx}
(i.e.\ up to the level of non-resonant contributions) if the virtuality of the resonances
is integrated over, as done in integrated cross sections or
most of the commonly used differential distributions.
For invariant-mass distributions of resonating particles,
non-factorizable corrections are non-vanishing, but turn out to be numerically
small as, e.g., discussed in the literature for single $\PW/\PZ$ production~\cite{Dittmaier:2014qza},
even to ${\cal O}(\alpha_{\mathrm{s}}\alpha)$,
or for the production of $\PW$-boson pairs~\cite{Beenakker:1997bp,Beenakker:1997ir,Denner:1997ia}
or $\PZ$-boson pairs~\cite{Denner:1998rh}.

The smallness of the sum of virtual and real non-factorizable corrections poses the
question about their relevance. Apart from the fact that there is no guarantee that
those effects are negligible unless they are calculated, the virtual non-factorizable corrections
alone represent an important building block in the ongoing effort of the
high-energy community in
automating NLO QCD and electroweak corrections to multi-particle processes.
On the side of real NLO corrections, the required evaluation of full LO
amplitudes, together with an appropriate subtraction of IR singularities, is under control
for up to $8{-}10$ final-state particles by automated systems such as
{\sc Sherpa}~\cite{Gleisberg:2008ta,Hoche:2010pf},
{\sc Madgraph}~\cite{Alwall:2007st,Frederix:2008hu},
or {\sc Helac-NLO}~\cite{Bevilacqua:2011xh,Czakon:2009ss}.
On the other hand, the much more complex evaluation of virtual one-loop amplitudes is confined to lower
multiplicities in spite of the great progress in recent years reached by the one-loop
matrix-element generators such as
{\sc BlackHat~\cite{Berger:2008sj}},
{\sc GoSam~\cite{Cullen:2011ac}},
{\sc Helac-NLO~\cite{Bevilacqua:2011xh}},
{\sc Madloop~\cite{Hirschi:2011pa}},
{\sc NJet~\cite{Badger:2012pg}},
{\sc OpenLoops~\cite{Cascioli:2011va}} and
{\sc Recola~\cite{Actis:2012qn}}.
A promising approach to drive automation to higher multiplicities in production processes
with several unstable particles in resonances---in particular in view of
electroweak corrections---is, thus, to make use of full matrix elements in LO
and on the side of the
real corrections, but to employ the PA for the virtual parts.
The factorizable virtual corrections can then be obtained with the above one-loop
matrix-element generators, accompanied by the non-factorizable virtual corrections,
for which we give explicit analytical results in this paper.
We note in passing that this kind of hybrid approach was already used in the
Monte Carlo generator {\sc RacoonWW}~\cite{Denner:1999gp,Denner:1999kn,Denner:2000bj,Denner:2002cg}
for $\PW$-pair production in $\Pep\Pem$ annihilation.

In detail, we present generic results on the non-factorizable virtual corrections
for the production of arbitrarily many resonances and their decays, i.e.\
we do not consider resonances that are part of cascade decays.
Moreover, we restrict our calculation to electroweak corrections to keep
the derivation and results transparent, but the modifications needed for
QCD corrections are straightforward.
Similar results were given in \citere{Accomando:2004de},
but without detailed derivation and somewhat less general.
Technically, pole expansions can be carried out on the basis of scattering amplitudes,
as done, e.g., in
Refs.~\cite{Melnikov:1995fx,Beenakker:1997bp,Beenakker:1997ir,Denner:1997ia,%
Dittmaier:2001ay,Accomando:2004de,Dittmaier:2014qza},
or alternatively with the help of specifically designed effective field theories, as
formulated in Refs.~\cite{Beneke:2003xh,Beneke:2004km}.
In this paper, we entirely analyze scattering amplitudes using
the Feynman-diagrammatic approach.

The paper is organized as follows:
In Sec.~\ref{sec:PA} we set our conventions and notations and
review the general structure of the pole approximation,
including the definition of factorizable and non-factorizable corrections.
Moreover, our strategy for calculating the non-factorizable corrections is
explained in detail there.
Section~\ref{sec:analytic-results} contains both our general results
and their illustration in applications to the Drell--Yan process,
to vector-boson pair production, and to vector-boson scattering.
Our conclusions are presented in Sec.~\ref{sec:conclusion}.
The appendices provide more details about the derivation
of our central results as well as supplementary formulas that are helpful
in the implementation of our results in computer codes.

\section{Pole approximation and non-factorizable corrections}
\label{sec:PA}

\subsection{Conventions and notations}
\label{sec:conventions-and-notations}

\begin{figure}
\centering
\includegraphics{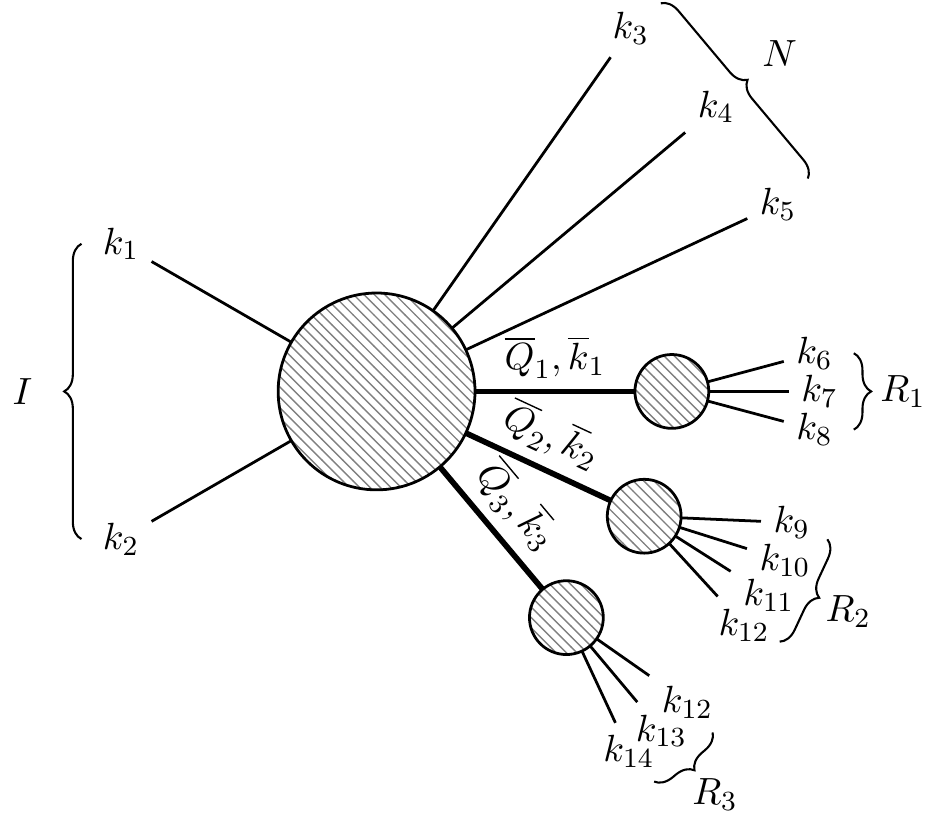}
\caption{Diagram for a typical process with multiple resonances illustrating the
labelling of external particles for a process $I \rightarrow F = N \cup R$.  The
particles with indices $i \in I$ are incoming, particles with indices $i \in F$
are outgoing. The outgoing particles either result from the decay of a resonant
particle, $i \in R$, or are directly produced without intermediate resonant
state, $i \in N$. There are $r$ resonances which have electric charges
$\overline{Q}_j$ and momenta $\overline{k}_j$ with $j \in
\overline{R} = \{ 1, \ldots, r \}$.
The decay products of resonance $j$ are labelled with $i \in R_j$.}
\label{fig:typical-process}
\end{figure}

Our conventions for labelling particles and momenta are illustrated in
Fig.~\ref{fig:typical-process}. We distinguish between initial- and final-state
particles where a final-state particle is either one of the $n$ non-resonant particles 
or a decay product of one of the $r$ resonant intermediate states.

We define the index set $I$ comprising the indices of incoming particles, the
$r$ sets $R_j$ containing the indices of the decay products of resonance $j$, the
set $\overline{R}$ of all $r$ resonances, and finally the set $N$ collecting the
$n$ remaining particles. Typically we have $I = \{ 1,2 \}$ and therefore $|I| =
2$, although we are not limited to this case. In summary, the numbers for
resonant and non-resonant particles are related to the index sets by
\begin{subequations}
\begin{align}
|R_j| &= r_j, \quad j \in \overline{R} \equiv \{ 1, \ldots, r \}, \\
|N| &= n.
\end{align}
\end{subequations}
For convenience we define
\begin{equation}
F = N \cup R, \quad R = \bigcup_{j=1}^r R_j \text{,}
\end{equation}
i.e.\ $F$ is the index set of all outgoing particles. The momentum of external
particle $i$ is labelled with $k_i$ for $i \in I \cup F$, where momenta are
defined to be outgoing. Incoming particles with incoming momenta $p_i, i \in I$,
therefore have momentum $p_i = - k_i$. The resonant particle $j$ has momentum
\begin{equation}
\overline{k}_j = \sum_{i \in R_j} k_i \text{.}
\end{equation}
We define invariants in the following way,
\begin{subequations}
\begin{alignat}{3}
s &= \Bigl( \sum_{i \in I} p_i \Bigr)^2 \text{,} \\
s_{ij} &= (k_i + k_j)^2 \text{,} &&& \quad i,j &\in I \cup F \text{,} \\
\overline{s}_{ij} &= (\overline{k}_i + k_j)^2 & \qquad i &\in \overline{R}
\text{,} & \quad j &\in I \cup F \text{,} \\
\tilde{s}_{ij} &= (\overline{k}_i - k_j)^2 \text{,} & \qquad i &\in \overline{R}
\text{,} & \quad j &\in I \cup F \text{,} \\
\overline{\overline{s}}_{ij} &= (\overline{k}_i + \overline{k}_j)^2 \text{,}
&\quad i,j &\in \overline{R} \text{,}
\end{alignat}
\end{subequations}
where whenever a quantity possesses a ``bar'' or a ``tilde'', it concerns a
resonant (intermediate) particle. The asymmetric sign convention in the
definition of $\overline{s}_{ij}$ and $\tilde{s}_{ij}$ accounts for the fact
that the momenta of the resonances are outgoing/incoming in the production/decay
subprocesses. The squared masses of the particles are
\begin{subequations}
\begin{align}
k_i^2 &= m_i^2 \text{,} \\
\overline{M}_j^2 &= M_j^2 - \mathrm{i} M_j \Gamma_j \text{,}
\end{align}
\end{subequations}
where $M_j$ and $\Gamma_j$ are the real mass and width parameters of the
unstable particle $j$. The final-state particles are taken to be massive, so
that potential collinear singularities in the case of light particles are
regularized by a small mass $m_i$. We also define the inverse propagator
denominators with complex masses for the resonant particle $j$ as
\begin{equation}
K_j = \overline{k}_j^2 - \overline{M}_j^2 \text{,} \quad j \in \overline{R}
\text{.}
\end{equation}
In order to regularize soft IR divergences we use an infinitesimal photon mass
$m_\gamma\to0$ and give a simple substitution rule to translate our results to
dimensional regularization in App.~\ref{app:scalar-integrals}.

Finally, each particle possesses an electric (relative) charge $Q_i$ so that
global charge conservation reads
\begin{equation}
\sum_{i \in I \cup F} Q_i \sigma_i = 0 \label{eq:global-charge-conservation}
\text{,}
\end{equation}
with sign factors $\sigma_i$ that are positive, $\sigma_i = +1$, for incoming
particles and outgoing antiparticles, and negative, $\sigma_i = -1$, for
incoming antiparticles and outgoing particles. Local charge conservation for the
resonance $j$ and its decay products reads
\begin{equation}
\overline{Q}_j = - \sum_{i \in R_j} Q_i \sigma_i \text{,} \quad \sum_{j=1}^r
\overline{Q}_j = \sum_{i \in I \cup N} \sigma_i Q_i
\label{eq:local-charge-conservation} \text{,}
\end{equation}
where $\overline{Q}_j$ is the electric charge of the produced resonance~$j$ of
particle or antiparticle type alike.

\subsection{Structure of the pole approximation}

\subsubsection{Factorizable corrections}

We define the LO matrix element in PA, $\mathcal{M}_\text{LO,PA}$,
as the product of the matrix elements for the
production of $r$ resonances with $n$ additional non-resonant states $N$,
$\mathcal{M}_\text{LO}^{I \rightarrow N,R}$, and the matrix elements of the
decays of each resonance $j$, $\mathcal{M}_\text{LO}^{j \rightarrow R_j}$,
multiplied by a product of $r$ propagators of the resonant particles,
\begin{equation}
\mathcal{M}_\text{LO,PA} = \sum_{\lambda_1, \ldots, \lambda_r} \left(
\prod_{i=1}^r \frac{1}{K_i} \right) \left[ \mathcal{M}_\text{LO}^{I \rightarrow
N,R} \left( \prod_{j=1}^r \mathcal{M}_\text{LO}^{j \rightarrow R_j} \right)
\right]_{\bigl\{ \overline{k}_l^2 \to \hat{\overline{k}}_l^2= M_l^2 \bigr\}_{l \in \overline{R}}} \text{.}
\label{eq:lo-matrix-element}
\end{equation}
This product involves the sum over the polarizations $\lambda_j$ of the
resonances, inducing spin correlations between the different production and
decay subprocesses. Note that the momenta of the resonances have to be set on
shell in the matrix elements of the subprocesses, $\mathcal{M}_\text{LO}^{I
\rightarrow N,R}$ and $\mathcal{M}_\text{LO}^{j \rightarrow R_j}$, otherwise the
constructed matrix element is not gauge invariant in general.  Since in PA we
only keep the leading contribution in the expansion about the resonance poles,
we can deform the original (off-shell) momenta $\overline{k}_i$ to momenta
$\hat{\overline{k}}_i$ that are on the (real) mass shell,
$\hat{\overline{k}}_l^2= M_l^2$,
which avoids the
unpleasant appearance of complex momentum variables.  The matrix element
$\mathcal{M}_\text{LO,PA}$ is, thus, the leading contribution of an expansion of
the full matrix element of the process $I \rightarrow F$ in the limit $\Gamma_j
\rightarrow 0$, 
where the widths $\Gamma_j$ in the denominators $1/K_j$ are kept.

We emphasize that the LO matrix element in PA, $\mathcal{M}_\text{LO,PA}$, is
only an auxiliary quantity in NLO predictions in PA, while LO cross sections
should be calculated with full LO matrix elements. Using
$\mathcal{M}_\text{LO,PA}$, e.g., in production processes of electroweak gauge
bosons $V=\PW,\PZ$ would neglect already terms of relative order ${\cal
O}(\Gamma_V/M_V)={\cal O}(\alpha)$, which is of the generic order of NLO
electroweak corrections.

The \emph{factorizable corrections} by definition comprise all corrections to
the various production and decay subprocesses, i.e.\ the corresponding matrix
element $\mathcal{M}_\text{virt,fact}$ is a sum of $r+1$ terms resulting from
$\mathcal{M}_\text{LO,PA}$ upon replacing one of the LO parts on the r.h.s.\ of
Eq.~\eqref{eq:lo-matrix-element} by the corresponding one-loop-corrected matrix
element $\mathcal{M}_\text{virt}$,
\begin{equation}
\begin{split}
\mathcal{M}_\text{virt,fact,PA} = \sum_{\lambda_1, \ldots, \lambda_r} \left(
\prod_{i=1}^r \frac{1}{K_i} \right) \Biggl[ &\mathcal{M}_\text{virt}^{I
\rightarrow N,R} \prod_{j=1}^r \mathcal{M}_\text{LO}^{j \rightarrow R_j} \\
+ &\mathcal{M}_\text{LO}^{I \rightarrow N,R} \sum_{k=1}^r
\mathcal{M}_\text{virt}^{k \rightarrow R_k} \prod_{j \neq k}^r
\mathcal{M}_\text{LO}^{j \rightarrow R_j} \Biggr]_{\bigl\{ \overline{k}_l^2
\to \hat{\overline{k}}_l^2= M_l^2 \bigr\}_{l \in \overline{R}}} \text{.}
\label{eq:virtual-factorizable-correction}
\end{split}
\end{equation}

\subsubsection{On-shell projection}
\label{sec:on-shell-projection}

Two different types of momenta enter Eqs.~\eqref{eq:lo-matrix-element} and
\eqref{eq:virtual-factorizable-correction}. The phase-space integral of the
corresponding cross-section contribution usually is based on the full phase
space determined by the momenta $k_a$ of all final-state particles $a \in F$,
where the intermediate momenta $\overline{k}_i$ are off their mass shell. These
are the momenta entering the propagator factor $\prod_i (1/K_i)$, while the
partial matrix elements appearing in the square brackets are parametrized by
{\it on-shell-projected momenta} $\hat k_a$ that result from all $k_a$ by some
deformation
\begin{equation}
\bigl\{ k_i \bigr\}_{i \in I \cup F} \;\to\; \bigl\{ \hat{k}_i \bigr\}_{i
\in I \cup F}
\end{equation}
in order to project the virtualities $\overline{k}_i^2$ of all resonances to
their real mass shells at $M_i^2$, i.e.\
\begin{equation}
\hat{\overline{k}}_i = \sum_{a \in R_i} \hat{k}_a \text{,} \qquad
\hat{\overline{k}}_i^2 = M_i^2 \text{,} \qquad
i\in\overline{R} \text{.}
\label{eq:on-shell-condition}
\end{equation}
This on-shell projection has to respect overall momentum conservation and all
mass-shell relations $k_a^2=\hat{k}_a^2=m_a^2$.
Note that the projection involves some
freedom, but the differences resulting from different definitions are of the
order of the otherwise neglected non-resonant contributions.

We suggest the following on-shell projection (which is a generalization of the
projection defined in Ref.~\cite{Denner:2000bj}) for our considered class of processes
with $r \ge 2$ resonances and possibly additional non-resonant particles
in the final state.
The on-shell projection preserves the momenta of the
initial-state and non-resonant final-state particles, i.e.\
\begin{equation}
\hat{k}_a = k_a \qquad a \in I \cup N \text{.}
\label{eq:non-resonant-on-shell-momenta}
\end{equation}
We construct the on-shell-projected momenta by selecting pairs of $i, j$
of resonances whose new momenta $\hat{k}_i$ and $\hat{k}_j$
are defined in
their centre-of-mass frame, i.e.\ in the frame where
$\vec{\overline{k}}_i+\vec{\overline{k}}_j=\sum_{a \in R_i \cup R_j}
\vec{k}_a = 0$. In this frame the momenta of the two resonances are back-to-back
and the velocities fixed by momentum conservation. We choose the direction of
the on-shell-projected momentum $\hat{\vec{\overline{k}}}_i$ of resonance~$i$
along its original direction $\vec{e}_i = \vec{\overline{k}}_i /
| \vec{\overline{k}}_i |$, which determines the on-shell-projected momenta as
follows,
\begin{subequations}\label{eq:on-shell-resonances}%
\begin{alignat}{2}
\hat{\overline{k}}_i^0 &= \frac{\overline{\overline{s}}_{ij} + M_i^2 - M_j^2}{2
\sqrt{\overline{\overline{s}}_{ij}}} \text{,} & \qquad
\hat{\vec{\overline{k}}}_i &= \frac{\sqrt{\lambda (\overline{\overline{s}}_{ij},
M_i^2, M_j^2)}}{2 \sqrt{\overline{\overline{s}}_{ij}}} \vec{e}_i \text{,} \\
\hat{\overline{k}}_j^0 &= \frac{\overline{\overline{s}}_{ij} - M_i^2 + M_j^2}{2
\sqrt{\overline{\overline{s}}_{ij}}} \text{,} & \qquad
\hat{\vec{\overline{k}}}_j &= - \hat{\vec{\overline{k}}}_i \text{,}
\label{eq:on-shell-momentum-conservation}
\end{alignat}
\end{subequations}
where $\lambda (x, y, z) = x^2 + y^2 + z^2 - 2xy - 2xz - 2yz$ is the well-known
triangle function.
Note that
this procedure leaves the sum of the two resonance four-momenta
(and thus also their invariant mass $\overline{\overline{s}}_{ij}$)
unchanged,
$\hat{\overline{k}}_i+ \hat{\overline{k}}_j= {\overline{k}}_i+ {\overline{k}}_j$.
Carrying out the procedure for all pairs of resonances
in $\overline{R}$ completes the on-shell projection if their
total number~$r$ is even. If there is an odd number of resonances, the remaining
resonance is paired with an already projected resonance momentum (preferably one
of Eq.~\eqref{eq:on-shell-momentum-conservation} where we did not preserve the
direction) and repeat the procedure for this pair once again.

The on-shell projection of the decay products of each resonance
can be done in a second step after fixing the resonance momenta
$\hat{\overline{k}}_i$ as above.
For simplicity we restrict ourselves to the case where a resonance~$i$
undergoes a $1 \to 2$ particle decay.
Denoting the two decay particles of $i$ by $a$ and $b$,
i.e.\ $R_i=\{ a, b \}$, we define the new momenta
$\hat k_a$ and $\hat k_b$ in the centre-of-mass frame of $\hat{\vec{\overline{k}}}_i$
as
\begin{subequations}\label{eq:on-shell-final-states}%
\begin{alignat}{2}
\hat{k}_a^0 &= \frac{M_i^2 + m_a^2 - m_b^2}{2 M_i} \text{,} & \qquad
\hat{\vec{k}}_a &= \frac{\sqrt{\lambda (M_i^2, m_a^2, m_b^2)}}{2 M_i} \vec{e}_a
\text{,} \\
\hat{k}_b^0 &= \frac{M_i^2 - m_a^2 + m_b^2}{2 M_i} \text{,} & \qquad
\hat{\vec{k}}_b &= - \hat{\vec{k}}_a \text{,}
\end{alignat}
\end{subequations}
where $\vec{e}_a = \vec{\overline{k}}_a / | \vec{\overline{k}}_a |$
is the direction of the original momentum $\vec{\overline{k}}_a$ in the
centre-of-mass frame of $\hat{\vec{\overline{k}}}_i$.
Note that this transformation is a simple rescaling of
$k_a$ and $k_b$ if $a$ and $b$ are massless.

For processes with a single resonance it is not possible to leave all of the
initial-state and non-resonant final-state momenta unmodified. In
Sec.~\ref{sec:single-z-or-w-boson-production} we give a suitable on-shell
projection for the case of no additional non-resonant particles and one
resonance.

\subsubsection{Non-factorizable corrections}
\label{sec:definition-of-the-non-factorizable-corrections}

Following the guideline of \citere{Denner:1997ia},
we define the non-factorizable virtual correction as the difference between the
full matrix element $\mathcal{M}_\text{virt}$ and the factorizable part in the
PA, i.e.\
\begin{equation}
\mathcal{M}_\text{virt,nfact,PA} \equiv \Bigl[ \mathcal{M}_\text{virt} -
\mathcal{M}_\text{virt,fact,PA} \Bigr]_{\text{res}, \bigl\{
\overline{k}_l^2,\overline{M}_l^2 \to M_l^2 \bigr\}_{l \in \overline{R}}}
\text{,} \label{eq:virtual-nonfactorizable-correction}
\end{equation}
where the subscript `res' indicates that after performing the loop integration
we keep only the resonant part of the expression. The additional subscript
$\bigl\{ \overline{k}_l^2,\overline{M}_l^2 \to M_l^2 \bigr\}_{l \in
\overline{R}}$ means that we set the virtualities and the complex masses of the
resonances to their real mass shell whenever possible, i.e.\ when the
replacements $\overline{k}_l^2 \to M_l^2$ and $\Gamma_l\to0$ do not lead to
singularities.  Apart from the terms where $\overline{k}_l^2$ and
$\overline{M}_l^2$ have to be kept, the non-factorizable matrix elements should
be evaluated with on-shell projected momenta to produce a well-defined result.

The procedure for deriving $\mathcal{M}_\text{virt,nfact,PA}$ will
be worked out in detail in
Sec.~\ref{sec:derivation-of-the-non-factorizable-corrections} below.
Here we just anticipate some basic features.
In contrast to the factorizable parts the non-factorizable corrections receive
contributions from diagrams in which the loop involves both production and decay
of the resonances, so that the expression does no longer factor in the
simple form of Eq.~\eqref{eq:virtual-factorizable-correction}, justifying the
name non-factorizable. However, as we will show in
Sec.~\ref{sec:derivation-of-the-non-factorizable-corrections}, the
non-factorizable corrections can be written as
\begin{equation}
2 \Re \left\{ \mathcal{M}_\text{LO,PA}^* \mathcal{M}_\text{virt,nfact,PA}
\right\} \equiv \left| \mathcal{M}_\text{LO,PA} \right|^2 \delta_\text{nfact}
\text{,} \label{eq:virt-nfact-PA}
\end{equation}
which defines the relative correction factor $\delta_\text{nfact}$, for which we
give an analytic expression in Sec.~\ref{sec:analytic-results}. In order to keep
the derivation and the final results transparent, in this paper we restrict
ourselves to the case of electroweak corrections, where only photon exchange
turns out to be relevant.

\subsection{Calculation of the non-factorizable corrections}
\label{sec:derivation-of-the-non-factorizable-corrections}

\subsubsection{Relevant Feynman diagrams}
\label{sec:relevant-feynman-diagrams}

In Eq.~\eqref{eq:virtual-nonfactorizable-correction} we have defined the
non-factorizable corrections as the resonant parts of the difference between the
full one-loop matrix elements and the factorizable terms. Thus, by definition
the sum of the factorizable and non-factorizable corrections, defined in
Eqs.~\eqref{eq:virtual-factorizable-correction} and
\eqref{eq:virtual-nonfactorizable-correction}, captures the full virtual
correction in PA.

\begin{figure}
\centering
\subfloat[$\mathrm{ff}'$]{\includegraphics{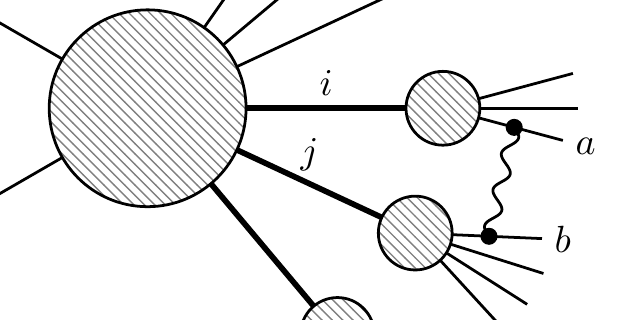}%
\label{fig:ffprime}}%
\subfloat[$\mathrm{nf}$]{\includegraphics{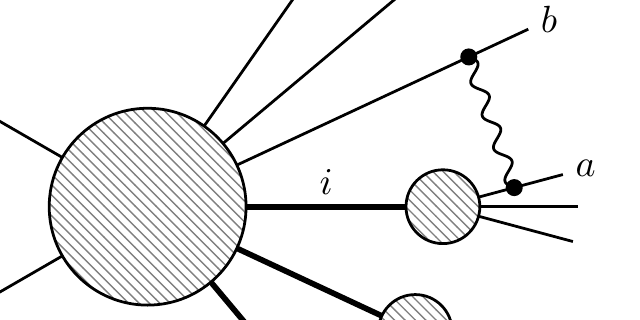}%
\label{fig:nf}} \\
\subfloat[$\mathrm{if}$]{\includegraphics{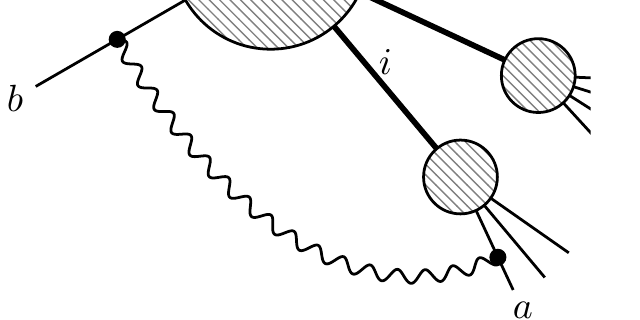}\label{fig:if}}%
\subfloat[$\mathrm{mf}'$]{\includegraphics{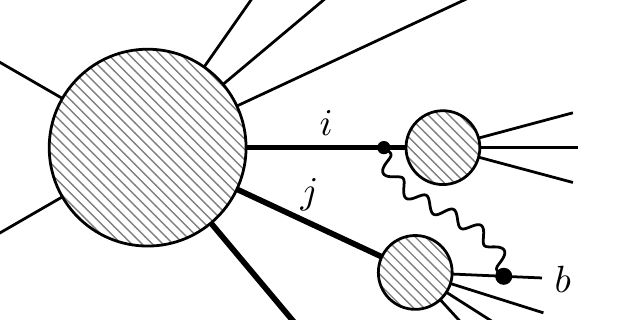}%
\label{fig:mfprime}} \\
\subfloat[$\mathrm{mf}$]{\includegraphics{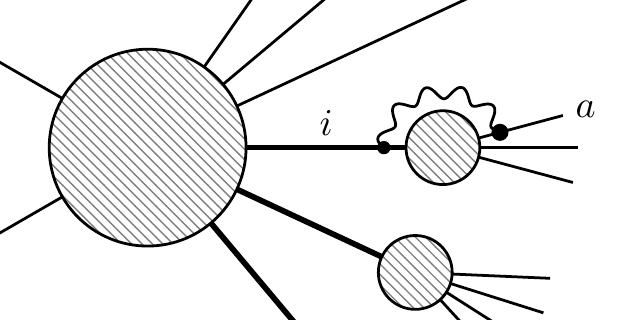}\label{fig:mf}}%
\subfloat[$\mathrm{mn}$]{\includegraphics{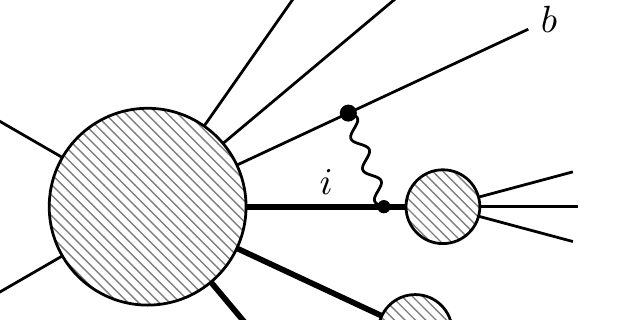}\label{fig:mn}} \\
\subfloat[$\mathrm{im}$]{\includegraphics{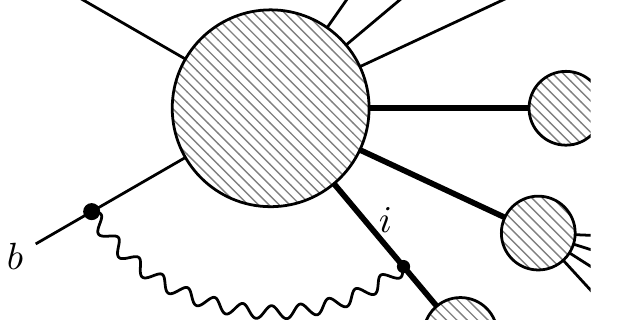}\label{fig:im}}%
\subfloat[$\mathrm{mm}'$]{\includegraphics{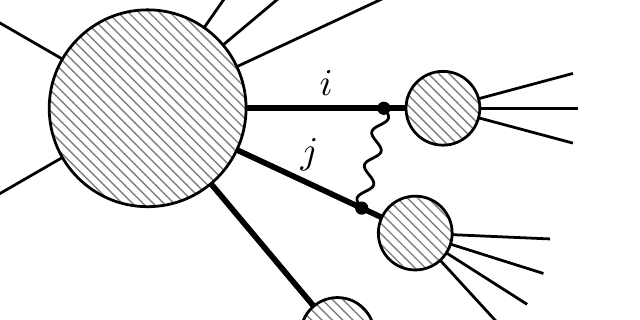}%
\label{fig:mmprime}} \\
\subfloat[$\mathrm{mm}$]{\includegraphics{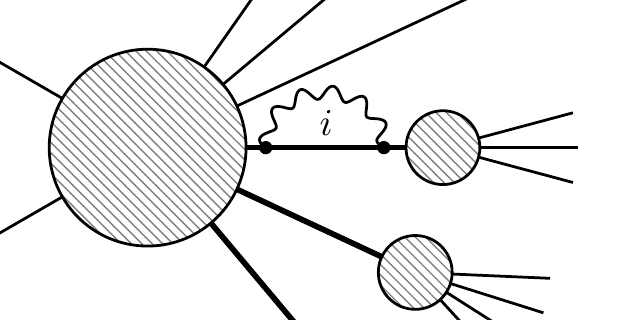}\label{fig:mm}}%
\caption{Feynman diagrams that contribute to the non-factorizable photonic
corrections. The diagrams \ref{fig:ffprime}, \ref{fig:nf}, \ref{fig:if}, and
\ref{fig:mfprime} are called {\it manifestly non-factorizable}, since they
do not contain factorizable contributions. They are type~(A) diagrams, as
defined in Sec.~\ref{sec:relevant-feynman-diagrams}. The remaining diagrams also
have factorizable parts and thus are of type~(B).}
\label{fig:non-factorizable-diagrams}
\end{figure}

Although the definition of the non-factorizable corrections involves the full
matrix element $\mathcal{M}_\text{virt}$, we do not need to know the full
expression of $\mathcal{M}_\text{virt}$, since only a specific set of diagrams
contributes to the non-factorizable parts. Following the arguments of
Refs.~\cite{Fadin:1993dz,Melnikov:1995fx,Beenakker:1997bp,Beenakker:1997ir,%
Denner:1997ia}, this set is identified as follows:

\begin{enumerate}
\item
By definition, all diagrams that do not involve the resonance pattern of the
considered process do not contribute to the resonant (factorizable or
non-factorizable) corrections. Since resonance factors may also emerge from the
loop integration, propagators in loops have to be included in the identification
of potential resonances. In a first step, certainly all diagrams can be omitted
that do not involve all relevant resonance propagators after omitting an
internal line in the loop. After this step, we are left with two types of
diagrams:
\begin{itemize}
\item[(A)] Diagrams in which at least one resonance
propagator~$j\in\overline{R}$ is confined in the loop. These are called
\textit{manifestly non-factorizable}.
\item[(B)] Diagrams in which all resonance propagators appear at least on one
tree-like line.
\end{itemize}
\item
Among the diagrams of type~(A) only those can develop a resonance corresponding
to the propagator~$j$ that is confined in a loop
if the loop contains a virtual photon
exchanged between external particles and/or resonances of the process,
because only then a soft IR divergence 
emerges.%
\footnote{The exchange of a
massless (or light) fermion does not produce the needed enhancement because of
the additional momentum term $\slashed{q}$ in the propagator numerator. Massless
or (light) scalars are ignored in this argument, since they are not part of the
SM or of any favoured extension.} This can be seen via simple power-counting in
momentum space.  Denoting the loop momentum on the propagator~$j$ by
$\overline{k}_j+q$, the resonance factor
$1/[(\overline{k}_j+q)^2-\overline{M}_j^2]$ receives support in the loop
integration only within a phase-space volume in which each component of $q$ is
of ${\cal O}(|\overline{k}_j^2-\overline{M}_j^2|/M_j)\sim {\cal O}(\Gamma_j)$.
To compensate this suppression factor $\propto\Gamma_j^4$ in the
four-dimensional loop integration, four powers of enhancement in the small
momentum $q$ are necessary. The only way to achieve this in a one-loop integral
is a soft divergence by a photon exchange (or a gluon in the QCD case), a
situation that can appear in two different ways. Firstly, the photon can be
exchanged between two different external particles $a$ and $b$, where the IR
divergence is produced by the factor $1/[(q^2-m_\gamma^2)(q^2+2k_a q)(q^2-2k_b
q)]$ composed of the three additional propagators. Secondly, the photon can be
exchanged between an external particle $a$ and another resonance $i \neq j$,
where the IR divergence is produced
by the factor%
\footnote{The factor $1/(q^2-2 \overline{k}_i q)$ actually results from a
decomposition of photon radiation off~$i$ into parts corresponding
to production and decay of resonance~$i$, which is achieved via a partial fractioning 
of propagators as shown in Eqs.~\eqref{eq:partfract1} and \eqref{eq:partfract2} below.
Without this decomposition this factor reads
$1/[(\overline{k}_i-q)^2-\overline{M}_i^2]$, i.e.\ the enhancement necessary in the
power-counting argument exists for small $q\sim{\cal O}(\Gamma_i)$.}
$1/[q^2-m_\gamma^2)(q^2+2 k_a q)(q^2-2 \overline{k}_i q)]$.
\item
The diagrams of type~(B) already contribute to the factorizable corrections,
because the respective loop subdiagrams contribute to an irreducible vertex
function that can be attributed to the production or one of the decay
subprocesses. Their factorizable contributions are obtained upon setting all
momenta $\overline{k}_i$ ($i\in\overline{R}$) of the resonances to their mass
shell everywhere but in the explicit propagator factors $1/K_i$.
Since we are only interested in the leading contribution of the expansion
about the resonances, we can neglect the decay widths $\Gamma_i$ when setting
$\overline{k}_i$ on shell, i.e.\ we can keep $\overline{k}_i^2=M_i^2$ real,
which conceptually and technically simplifies the evaluation of the factorizable
corrections significantly. Diagrams of type~(B)
can, thus, only contribute to
the non-factorizable corrections if the two steps of the loop integration and
the transition to $\overline{k}_i^2=M_i^2$ in the loop do not
commute.\footnote{In the full contribution the loop integration is done first,
followed by the identification of the resonant parts upon taking
$\overline{k}_i^2\to M_i^2$. In the factorizable contributions
$\overline{k}_i^2=M_i^2$ is set in the integrand before the loop integration.}
This can only happen if the process of setting $\overline{k}_i^2\to M_i^2$
before the loop integration leads to a singularity for at least one resonance,
which in turn is only the case if the loop contains a photon exchanged between
resonance $i$ and an external particle or another resonance.
\end{enumerate}
In summary, non-factorizable corrections are due to diagrams that result from
the corresponding LO diagrams by allowing for photon exchange between external
particles of different subprocesses and
resonances in all possible ways.
The corresponding generic Feynman diagrams are illustrated in
Fig.~\ref{fig:non-factorizable-diagrams}.

\subsubsection{Extended soft-photon approximation}

Considering the diagrams with non-factorizable contributions in more detail in
momentum space, only loop momenta $q$ of the internal photon with components of
${\cal O}(\Gamma_i)$ can contribute to the non-factorizable corrections, where
$\Gamma_i$ generically stands for the energy scale determined by the decay
widths of the resonances. For diagrams of type~(A) this is obvious, for diagrams
of type~(B) this is a consequence of the fact that the difference between the
full diagram and its factorizable part can only develop a resonant part for such
small $q$. This observation is the basis for the evaluation of the
non-factorizable contributions in ``extended soft-photon approximation'' (ESPA)
which is a modification of the commonly used ``soft-photon approximation'',
which is based on the eikonal currents of soft photons. The modification
concerns the fact that the soft momentum $q$ is kept in the denominators of the
resonance propagators, but are neglected elsewhere as usual. In particular, $q$
can be set to zero in the numerator of Feynman diagrams and in the denominators
of all propagators that do not contribute to the soft divergences mentioned
above. As a consequence, the non-factorizable corrections can be deduced from
scalar one-loop integrals (i.e.\ without integration momenta in the numerator)
with at most five propagators in the loop integration
(largest number of loop propagators in Fig.~\ref{fig:non-factorizable-diagrams}),
and the resulting
correction factorizes from the underlying LO diagram, as already anticipated in
Eq.~\eqref{eq:virt-nfact-PA}.

Now we are able to start with the generic construction of the non-factorizable
contributions within the ESPA. The coupling of the soft photon to an external
particle, either incoming or outgoing, within the ESPA is exactly the same as in
the usual eikonal approximation, i.e.\ coupling the photon with outgoing
momentum~$q$ to the external line $a$ with momentum $k_a$ and electric charge
$Q_a$ modifies the underlying LO amplitude by the eikonal current factor
\begin{equation}
j^\mu_{\eik,a}(q) = - \frac{2e \sigma_a Q_a k_a^\mu}{q^2+2qk_a},
\end{equation}
where $a$ can be incoming or outgoing with the sign $\sigma_a=\pm1$ as defined
before, but $k_a$ is formally outgoing. Here and in the following, the $q^2$
term in a propagator denominator is always implicitly understood to contain
Feynman's $\ieps$ prescription according to $q^2+\ieps$. The usual soft-photon
approximation combines the individual contributions to the eikonal currents to a
full eikonal current $J^\mu_{\eik}(q) = \sum_a j^\mu_{\eik,a}(q)$, where the sum
runs over all external particles $a$, and the soft-photon factor that multiplies
$|\mathcal{M}_\text{LO}|^2$ is proportional to the integral $\int\rd^D q\,
J_{\eik}(q) \cdot J_{\eik}(-q)/(q^2-m_\gamma^2)$.
We will generalize the eikonal
currents to ESPA currents upon including contributions from the resonances, so
that individual currents can be attributed to the production and decay
subprocesses, $J_{\pro}$ and $J_{\dec,i}$. The factorizable corrections will
then be identified with the diagonal contributions $J_{\pro}(q)\cdot
J_{\pro}(-q)$ and $J_{\dec,i}(q)\cdot J_{\dec,i}(-q)$, while the
non-factorizable corrections correspond to non-diagonal terms $J_{\pro}(q)\cdot
J_{\dec,i}(-q)$ and $J_{\dec,i}(q)\cdot J_{\dec,j}(-q)$, where the photon is
exchanged by different subprocesses.

We first define the contributions of external particles to the ESPA currents.
Taking into account that outgoing lines $a\in R_i$ always result from resonance
$i\in \overline{R}$, we include the modification of the resonance factor by the
photon momentum in the definition of the ESPA current factors,
\begin{subequations}
\begin{align}
j^\mu_{a}(q) & = j^\mu_{\eik,a}(q) \frac{K_i}{K_i(q)} = - \frac{2e \sigma_a Q_a
k_a^\mu}{q^2+2qk_a} \, \frac{K_i}{K_i(q)}, & a & \in R_i, \; i\in \overline{R},
\label{eq:ESPA-FSR} \\
j^\mu_{a}(q) & = j^\mu_{\eik,a}(q) = - \frac{2e \sigma_a Q_a
k_a^\mu}{q^2+2qk_a}, & a & \in I\cup N \text{,}
\end{align}
\end{subequations}
where
\begin{equation}
K_i (q) = (\overline{k}_i + q)^2 - \overline{M}_i^2 = q^2 + 2 q \overline{k}_i +
K_i \text{.}
\end{equation}

Photon radiation off a resonance $i\in \overline{R}$ can be described
by similar factors, but their derivation is somewhat more involved.
The first step in this derivation is to analyse the emission of
a soft photon with momentum $q$ off $i$, where the components of $q$
are of ${\cal O}(\Gamma_i)$.
In App.~\ref{app:soft-photon} we show for the relevant cases of resonances
with spin 0, 1/2, or 1 that
\begin{subequations}
\begin{align}
\vcenter{\hbox{\includegraphics{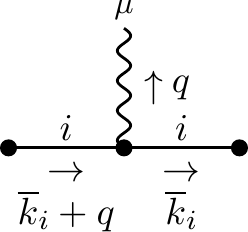}}} &\;\;= \;\;
\frac{2e \overline{Q}_i \overline{k}_i^\mu}{K_i(q)} \;\times\;
\vcenter{\hbox{\includegraphics{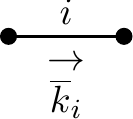}}} \;\;,
\label{eq:Rphoton1}
\\
\vcenter{\hbox{\includegraphics{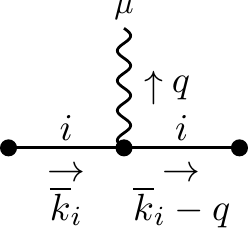}}} &\;\;= \;\;
\frac{2e \overline{Q}_i \overline{k}_i^\mu}{K_i(-q)} \;\times\;
\vcenter{\hbox{\includegraphics{feynman-diagrams/propagator1}}} \;\;,
\label{eq:Rphoton2}
\end{align}
\end{subequations}
where the graphically represented propagators on the right-hand sides
are proportional to $1/K_i$.
Here, the charge $\overline{Q}_i$ refers to a particle or antiparticle flowing from
the production part on the left to its decay part on the right.
The subdiagram on the l.h.s.\
of Eq.~\eqref{eq:Rphoton1} belongs to a graph in which resonance $i$
exchanges a photon with an external particle of the production part,
or with another resonance $j\ne i$,
or with any external particle of a final-state particle of any other
resonance $j\ne i$.
The second diagram belongs to the situation where the photon exchange
happens between resonance $i$ and one of its decay particles.
In both situations the photon emission off the resonance can be split
into an emission part before or after the resonant propagation
using the partial fractionings
\begin{subequations}
\begin{align}
\frac{1}{K_i(q)K_i} & = \frac{1}{q^2+2\overline{k}_i q}
\left[\frac{1}{K_i} - \frac{1}{K_i(q)}\right],
\label{eq:partfract1}
\\
\frac{1}{K_iK_i(-q)} & = \frac{1}{q^2-2\overline{k}_i q}
\left[\frac{1}{K_i} - \frac{1}{K_i(-q)}\right].
\label{eq:partfract2}
\end{align}
\end{subequations}
Applied to the two subgraphs of Eqs.~\eqref{eq:Rphoton1} and \eqref{eq:Rphoton2},
this leads to
\begin{subequations}
\begin{align}
\vcenter{\hbox{\includegraphics{feynman-diagrams/current1}}} &\;\;= \;\;
\frac{2e \overline{Q}_i \overline{k}_i^\mu}{q^2+2\overline{k}_i q} \; \times \;
\left[\;
\vcenter{\hbox{\includegraphics{feynman-diagrams/propagator1}}}
\;\;-\;\;
\vcenter{\hbox{\includegraphics{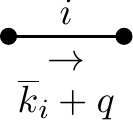}}}
\;\right],
\label{eq:Rphoton3}
\\
\vcenter{\hbox{\includegraphics{feynman-diagrams/current2}}} &\;\;= \;\;
\frac{2e \overline{Q}_i \overline{k}_i^\mu}{q^2-2\overline{k}_i q} \; \times \;
\left[\;
\vcenter{\hbox{\includegraphics{feynman-diagrams/propagator1}}}
\;\;-\;\;
\vcenter{\hbox{\includegraphics{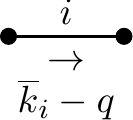}}}
\;\right].
\label{eq:Rphoton4}
\end{align}
\end{subequations}

The first contribution on the r.h.s.\ of Eq.~\eqref{eq:Rphoton3}, which is proportional to $1/K_i$,
corresponds to photon radiation during the production of the resonance.
We attribute the ESPA current
\begin{align}
\overline{j}^\mu_{\out,i}(q) =
\frac{2e \overline{Q}_i \overline{k}_i^\mu}{q^2+2\overline{k}_i q}
\end{align}
to an outgoing resonance $i$ when it exchanges a photon with any particle
of the production phase or the decay of any other resonance $j\ne i$.
Applying this current factor to the corresponding LO matrix element
describes soft-photon emission
off a resonance of particle or antiparticle type during its
production phase.
The second subdiagram on the r.h.s.\ of Eq.~\eqref{eq:Rphoton3}
corresponds to photon radiation during the decay of the resonance.
We define the ESPA current
\begin{align}
\overline{j}^\mu_{\inc,i}(q) =
- \frac{2e \overline{Q}_i \overline{k}_i^\mu}{q^2+2\overline{k}_i q} \,
\frac{K_i}{K_i(q)},
\end{align}
which describes soft-photon emission off the resonance~$i$
during its decay phase.
The factor $K_i/K_i(q)$ accounts for the fact that the propagator $1/K_i$
is included in the LO amplitude, but not $1/K_i(q)$.
Using these results in combination with the ESPA currents~\eqref{eq:ESPA-FSR}
describing radiation off the decay products, we can define the complete
ESPA current $J^\mu_{\dec,i}(q)$ for the decay of resonance~$i$,
\begin{equation}
J^\mu_{\dec,i}(q)
= \overline{j}^\mu_{\inc,i}(q) + \sum_{a\in R_i} j^\mu_{a}(q)
= \left[ - \frac{2e \overline{Q}_i \overline{k}_i^\mu}{q^2+2\overline{k}_i q}
- \sum_{a\in R_i} \frac{2e \sigma_a Q_a k_a^\mu}{q^2+2k_a q}
\right] \frac{K_i}{K_i(q)}.
\end{equation}
The combination $J_{\dec,i}(q)\cdot J_{\dec,j}(-q)$ is, thus, relevant
for the non-factorizable corrections induced by photon exchange between
the two decay subprocesses of two different resonances $i$ and $j$.

The second type of photon emission, treated in Eq.~\eqref{eq:Rphoton4},
is needed to describe photon exchange between a resonance~$i$ in its
production phase and itself or one of its decay products.
More precisely, it is the second term on the r.h.s.\ of Eq.~\eqref{eq:Rphoton4}
that corresponds to this situation, since the corresponding $i$~propagator
carries the momentum $\overline{k}_i-q$, where
the $i$~momentum is reduced by photon radiation.
For an outgoing resonance
we define the ESPA current
\begin{align}
\widetilde{j}^\mu_{\out,i}(q) =
- \frac{2e \overline{Q}_i \overline{k}_i^\mu}{q^2-2\overline{k}_i q},
\end{align}
where we have
not included the factor $K_i/K_i(-q)$ in the definition in order to avoid
double counting this factor, because it is already included in the
definition of $J^\mu_{\dec,i}(-q)$ which will multiply
$\widetilde{j}^\mu_{\out,i}(q)$ in the calculation of the corresponding photon-exchange
diagrams.
The full ESPA current $J^\mu_{\pro,i}(q)$
for the production of resonance~$i$ to describe photon
exchange with the decay subprocess of $i$, then consists of three different
types of contributions:
the first where the photon is attached to resonance~$i\in \overline{R}$,
the second where the photon is attached to any other
resonance~$j\in \overline{R}, j\ne i$, and
the third where the photon is attached to external stable
particles~$a \in I\cup N$, of the production phase,
\begin{align}
J^\mu_{\pro,i}(q) &\;=\; \widetilde{j}^\mu_{\out,i}(q)
+ \sum_{\substack{j\in\overline{R} \\ j\ne i}} \overline{j}^\mu_{\out,j}(q)
+ \sum_{a\in I\cup N} j^\mu_{a}(q)
\nonumber\\
&\;=\;
- \frac{2e \overline{Q}_i \overline{k}_i^\mu}{q^2-2\overline{k}_i q}
+ \sum_{\substack{j\in\overline{R} \\ j\ne i}}
\frac{2e \overline{Q}_j \overline{k}_j^\mu}{q^2+2\overline{k}_j q}
- \sum_{a\in I\cup N} \frac{2e \sigma_a Q_a k_a^\mu}{q^2+2k_a q}.
\end{align}
Since the first term in Eq.~\eqref{eq:Rphoton4} involves
the propagator factor $1/K_i$ without photon momentum, this contribution
corresponds to photon exchange between the resonance~$i$ during its decay phase
and any particle taking part in the decay of $i$. This term in
Eq.~\eqref{eq:Rphoton4} is, thus, only relevant for the factorizable
soft-photonic corrections to the decay of $i$.

In summary, the complete set of non-factorizable contributions
can be written as
\begin{equation}
\delta_\text{nfact} = 2\Re\biggl\{
(2\pi\mu)^{4-D}
\int \rd^D q \, \frac{-\ri}{q^2-m_\gamma^2} \,
\biggl[ \,
\sum_{i\in\overline{R}} J_{\pro,i}(q) \cdot J_{\dec,i}(-q)
+
\sum_{\substack{i,j\in\overline{R} \\ i\ne j}} J_{\dec,i}(q) \cdot J_{\dec,j}(-q)
\biggr] \biggr\} \text{,}
\label{eq:deltanfact}
\end{equation}
in $D$ dimensions in order to regularize occurring UV~divergences.
Setting $m_\gamma$ to zero, directly produces the result for $\delta_\text{nfact}$ 
where soft IR~divergences are regularized dimensionally. 
By construction, the correction factor
$\delta_\text{nfact}$ is a gauge-invariant quantity.
Its derivation starts from the
difference~\eqref{eq:virtual-nonfactorizable-correction}
of the full matrix element and the corresponding factorizable corrections,
which are both gauge invariant.
Picking then the resonant parts from this difference in a consistent way
and dividing it by $|\mathcal{M}_\text{LO,PA}|^2$, leads to a gauge-invariant
result.
Electromagnetic gauge invariance is also reflected by the ESPA
currents $J^\mu_{\pro,i}(q)$ and $J^\mu_{\dec,i}(q)$,
since contracting them with $q_\mu$ gives zero up to terms of
${\cal O}(q^2)$, which, however, do only influence non-resonant
contributions and are thus negligible.
Therefore, in principle the ESPA currents and the non-factorizable corrections
could be defined upon setting the $q^2$ terms to zero, as for instance done in
Refs.~\cite{Beenakker:1997bp,Beenakker:1997ir} for W-pair production. We have
decided to keep the $q^2$ terms, firstly to be able to make direct use of
standard scalar integrals, and secondly to avoid artificial ultraviolet
divergences in non-resonant contributions.

It should be noted that $r$ different ESPA currents $J^\mu_{\pro,i} (q)$ are
necessary
to correctly describe photon exchange
between the production of resonance $i$ and the decay subprocesses of
resonance $j$, since the momentum flows for $i=j$ and $i \neq j$ are not the
same.\footnote{This fact was already realized in the calculation of
non-factorizable corrections to $e^+ e^- \to \mathrm{W} \mathrm{W} \to 4 f$ in
Refs.~\cite{Beenakker:1997bp,Beenakker:1997ir}, but overlooked in the
(correct)
calculation of Ref.~\cite{Denner:1997ia} where currents were only introduced for
illustration.}

\section{Analytic results for the non-factorizable corrections}
\label{sec:analytic-results}

\subsection{Generic result}

Having derived Eq.~\eqref{eq:deltanfact}, it is straightforward to translate all
individual contributions to the correction factor $\delta_\text{nfact}$ shown in
Fig.~\ref{fig:non-factorizable-diagrams} into a form expressed in terms standard
scalar one-loop integrals. This task is carried out in detail in
App.~\ref{app:virtual-corrections}, and some of the relevant one-loop integrals
are collected in App.~\ref{app:scalar-integrals}. The explicit results are given
by
\begin{equation}
\begin{split}
\delta_\text{nfact} = - \sum_{i=1}^r \sum_{j=i+1}^r \sum_{a \in R_i} \sum_{b \in
R_j} &\sigma_a \sigma_b Q_a Q_b \frac{\alpha}{\pi} \Re \left\{ \Delta \right\}
\\
- \sum_{i=1}^r \sum_{a \in R_i} \sum_{b \in N \cup I} &\sigma_a \sigma_b Q_a Q_b
\frac{\alpha}{\pi} \Re \left\{ \Delta' \right\}
\end{split}
\label{eq:non-factorizable-contribution}
\end{equation}
with functions
\begin{subequations}
\begin{align}
\Delta (i,a; j,b) &= \Delta_\mathrm{mm} + \Delta_\mathrm{mf} +
\Delta_\mathrm{mm'} + \Delta_\mathrm{mf'} + \Delta_\mathrm{ff'} , \\
\Delta' (i,a;b) &= \Delta'_\mathrm{mm} + \Delta'_\mathrm{mf} +
\Delta_\mathrm{xf} + \Delta_\mathrm{xm} ,
\label{eq:Deltaprime}
\end{align}
\end{subequations}
which depend on the indices of the external particles $a,b$ and the resonances
$i,j$ to which they are connected. Some of the indices $i,j,a,b$ might be
omitted if they do not appear in the considered subcontribution. Depending on
whether the index $b$ refers to the initial or final state, the contribution
$\mathit{xf}$ denotes either $\mathit{if}$ or $\mathit{nf}$, and for the
contribution $\mathit{xm}$ either $\mathit{im}$ or $\mathit{mn}$.

The matrix elements for diagrams of the type $\mathit{mm}$ (Fig.~\ref{fig:mm})
and $\mathit{mf}$ (Fig.~\ref{fig:mf}) are proportional to $\overline{Q}_i^2$ and
$\overline{Q}_i Q_b$, respectively, so that we have used
global charge conservation,
Eq.~\eqref{eq:global-charge-conservation}, to fit them in the summation
structure of Eq.~\eqref{eq:non-factorizable-contribution}. This is also the
reason why their contributions appear in both $\Delta$ and $\Delta'$.
Furthermore the $\mathit{mf'}$ contribution 
appears twice because we sum over
$i<j$. The relations
between the functions $\Delta_{\cdots}$ and
$\Delta_{\cdots}'$ are
\begin{subequations}
\begin{align}
\Delta_\mathrm{mm} (i;j) &= \Delta'_\mathrm{mm} (i) + \Delta'_\mathrm{mm} (j)
\text{,} \\
\Delta_\mathrm{mf} (i,a;j,b) &= \Delta'_\mathrm{mf} (i,a) + \Delta'_\mathrm{mf}
(j,b) \text{,} \\
\Delta_\mathrm{mf'} (i,a;j,b) &= \Delta'_\mathrm{mf'} (i;j,b) +
\Delta'_\mathrm{mf'} (j;i,a) \text{,}
\end{align}\label{eq:delta-prime-relations}%
\end{subequations}
so that we only need to give the primed functions $\Delta'_{\cdots}$.

The virtual parts of the not manifestly non-factorizable contributions are
\begin{subequations}\label{eq:non-manifestly-nf-contributions}
\begin{align}
\Delta_\mathrm{mm'} &\sim
\begin{aligned}[t]
- \left( \overline{\overline{s}}_{ij} - M_i^2 - M_j^2 \right) \bigl\{ &C_0
\left( \overline{k}_i^2, \overline{\overline{s}}_{ij}, \overline{k}_j, 0,
\overline{M}_i^2, \overline{M}_j^2 \right) \\
&- C_0 \left( M_i^2, \overline{\overline{s}}_{ij}, M_j^2, m_\gamma^2, M_i^2,
M_j^2 \right) \bigr\} \text{,} \label{eq:mm-prime-contribution}
\end{aligned} \\
\Delta'_\mathrm{mm} &\sim 2 M_i^2 \left\{ \frac{B_0 \left( \overline{k}_i^2, 0,
\overline{M}_i^2 \right) - B_0 \left( \overline{M}_i^2, m_\gamma^2,
\overline{M}_i^2 \right)}{K_i} - B'_0 \left( M_i^2, m_\gamma^2, M_i^2 \right)
\right\} \text{,} \\
\Delta'_\mathrm{mf} &\sim
\begin{aligned}[t]
- (\tilde{s}_{ia} - M_i^2 - m_a^2) \bigl\{ &C_0 \left( \overline{k}_i^2,
\tilde{s}_{ia}, m_a^2, 0, \overline{M}_i^2, m_a^2 \right) \\
&- C_0 \left( M_i^2, \tilde{s}_{ia}, m_a^2, m_\gamma^2, M_i^2, m_a^2 \right)
\bigr\} \text{,}
\end{aligned} \\
\Delta_\mathrm{xm} &\sim
\begin{aligned}[t]
- \left( \overline{s}_{ib} - M_i^2 - m_b^2 \right) \bigl\{ &C_0 \left(
\overline{k}_i^2 , \overline{s}_{ib}, m_b^2, 0, \overline{M}_i^2, m_b^2 \right)
\\
&- C_0 \left( M_i^2, \overline{s}_{ib}, m_b^2, m_\gamma^2, M_i^2, m_b^2 \right)
\bigr\} \text{.}
\end{aligned}
\end{align}
\end{subequations}
The parts in curly brackets contain the subtraction of 
the respective factorizable parts which are obtained
by setting the virtualities of the resonances on their real masses before the
loop integration, as discussed in
Sec.~\ref{sec:definition-of-the-non-factorizable-corrections}.

The manifestly non-factorizable virtual contributions read
\begin{subequations}\label{eq:manifestly-nf-contributions}
\begin{align}
\Delta_\mathrm{ff'} &\sim - (s_{ab} - m_a^2 - m_b^2) K_i K_j E_0 \left( k_a,
\overline{k}_i, - \overline{k}_j, - k_b, m_\gamma^2, m_a^2, \overline{M}_i^2,
\overline{M}_j^2, m_b^2 \right) \text{,} \label{eq:ff-prime-contribution} \\
\Delta'_{\mathrm{mf}'} &\sim
- (\overline{s}_{ib} - M_i^2 - m_b^2) K_j 
D_0 \left( \overline{k}_i, -\overline{k}_j, -k_b, m_\gamma^2, \overline{M}_i^2, 
\overline{M}_j^2, m_b^2 \right)
\text{,} \label{eq:mf-prime-contribution} \\
\Delta_\mathrm{xf} &\sim - (s_{ab} - m_a^2 - m_b^2) K_i D_0 \left( k_a,
\overline{k}_i, - k_b, m_\gamma^2, m_a^2, \overline{M}_i^2, m_b^2 \right)
\text{.} \label{eq:xf-contribution}
\end{align}
\end{subequations}
These contributions do not have factorizable counterparts.

For the second sum in Eq.~\eqref{eq:non-factorizable-contribution} it is
instructive to write down an explicit expression, at least for the
case of massless external particles. Using the loop integrals given
in App.~\ref{app:scalar-integrals} for $m_a,m_b\to0$, 
the function $\Delta'$ reads
\begin{equation}
\Delta' \sim 2 \left[ 
\ln\left( \frac{M_i^2 - \widetilde{s}_{ia}}{M_i^2}\right)
+\ln\left(\frac{M_i^2 - \overline{s}_{ib}}{-s_{ab}} \right)
- 1 \right] 
\ln \left( \frac{- K_i}{m_\gamma M_i}
\right) + 
\cLi2\left( \frac{M_i^2 - \widetilde{s}_{ia}}{M_i^2} ,
\frac{M_i^2 - \overline{s}_{ib}}{-s_{ab}} \right) 
+ 2 \text{.} 
\label{eq:non-resonant-initial-final-contribution}
\end{equation}
where $s_{ab}$, $\widetilde{s}_{ia}$, and $\overline{s}_{ib}$ are implicitly understood
as $s_{ab}+\ri0$, $\widetilde{s}_{ia}+\ri0$, and $\overline{s}_{ib}+\ri0$, respectively.
The dilogarithmic function $\cLi2$ is defined in Eq.~\eqref{eq:cLi}.

The correction factor $\delta_\text{nfact}$ contains soft divergences which are
regularized as terms proportional to $\ln m_\gamma$ (or poles $1/\epsilon$ in $D
= 4 - 2 \epsilon$ dimensions, cf.\ App.~\ref{app:scalar-integrals}). These terms
always appear as logarithms $\ln [ m_\gamma M_i/(\overline{M}_i^2 -
\overline{k}_i^2) ]$ as a result of the connection between the soft divergence
in the loop integration and the resonance at $\overline{k}_i^2=M_i^2$.

Note, however, that the whole correction factor $\delta_\text{nfact}$ is free of
mass singularities of the external particles $a,b$ if one or more masses $m_a,
m_b$ become small. In the subcontribution of
Eq.~\eqref{eq:non-resonant-initial-final-contribution}, this is directly
visible, but for the other contributions there is a non-trivial cancellation
between the corresponding mass singularities that appear in individual
contributions. For small masses $m_a, m_b$ it is, thus, possible to set the masses
to zero consistently, which changes individual singular loop integrals, but not the
final result for $\delta_\text{nfact}$.
In view of the limits $K_i\to0$,
note that there are two different types of non-analytic terms: The
mentioned $\ln K_i$ terms and rational functions of the form $K_i K_j / (a K_i^2
+ b K_i K_j + c K_j^2)$ originating from the five-point functions of
Eq.~\eqref{eq:ff-prime-contribution}, where $a, b, c$ are polynomial in
kinematical invariants. Terms of the latter type require at least two different
resonances and already appeared in the treatment of the W-pair
production~\cite{Beenakker:1997bp,Beenakker:1997ir,Denner:1997ia,%
Accomando:2004de,Billoni:2013aba,Bredenstein:2005zk}.

In order not to spoil the cancellation of mass singularities, it is essential to
use a unique procedure to isolate the non-analytic terms in the limit
$\overline{K}_i^2,\overline{M}_i^2\to M_i^2$ and to perform the on-shell
projection of the phase space in the regular terms.

Our results on photonic non-factorizable corrections 
confirm the generic results given in the appendix of
Ref.~\cite{Accomando:2004de}, which were formulated for
several resonances and non-resonant final-state particles as well,
though without details on their derivation.
The specific formulas of Ref.~\cite{Accomando:2004de} are given for the
situation where resonances decay into two massless particles, an 
assumption we do not make.
Moreover, we have presented a detailed general derivation of the
photonic non-factorizable corrections, including a definition of
the underlying ESPA current.

\subsection{Examples}

\subsubsection{Single Z- or W-boson production in hadronic collisions}
\label{sec:single-z-or-w-boson-production}

The simplest application of Eq.~\eqref{eq:non-factorizable-contribution} is the
production of a single resonance, e.g.\ the Drell--Yan-like production $q
\overline{q} \rightarrow \mathrm{Z} \rightarrow \ell^- \ell^+$ or $q
\overline{q}' \rightarrow \mathrm{W}^\pm \rightarrow \nu_\ell \ell^+ /
\ell^-\overline{\nu}_\ell$. 
There is only one resonance ($r=i=1$) and no
additional non-resonant particles in the final state ($n=0$),
so that Eq.~\eqref{eq:non-factorizable-contribution}
simplifies to
\begin{equation}
\delta_\text{nfact} = - \sum_{a \in R_1} \sum_{b \in I} \sigma_a Q_a \sigma_b
Q_b \, \frac{\alpha}{\pi} \Re \left\{ \Delta' \right\} \text{.}
\label{eq:single-resonance-nfact}
\end{equation}
Since the external fermion masses are negligible, we can make use of
$\Delta'$ as given in Eq.~\eqref{eq:non-resonant-initial-final-contribution}.
The relevant kinematical invariants read
\begin{equation}
\begin{alignedat}{2}
s_{12} = s_{34} &= 2 k_1 \cdot k_2 = \hphantom{-} 2 p_1 \cdot p_2 = s \text{,}
\qquad \overline{s}_{11} &= \overline{s}_{12} &= 0 \text{,} \\
s_{13} = s_{24} &= 2 k_1 \cdot k_3 = - 2 p_1 \cdot k_3 = t \text{,} \qquad
\widetilde{s}_{13} &= \widetilde{s}_{14} &= 0 \text{,} \\
s_{14} = s_{23} &= 2 k_1 \cdot k_4 = - 2 p_1 \cdot k_4 = u \text{,}
\end{alignedat}
\end{equation}
where we have taken the numbering $I=\{1,2\}$, $R_1=\{3,4\}$ and
$s,t,u$ are the usual Mandelstam variables.
With the particle ordering defined above, the sign factors $\sigma_a$ are
\begin{equation}
\sigma_1 = -\sigma_2 = 1, \quad
\sigma_3 = -\sigma_4 = -1.
\end{equation}

For the case of $\PWpm$ production/decay, the charge assignment is
\begin{equation}
Q_1 = Q_\Pu, \quad
Q_2 = Q_\Pd, \quad
Q_3 = Q_\nu = 0, \quad
Q_4 = Q_\ell = -1, \quad
\overline{Q}_1 = Q_\Pu - Q_\Pd,
\end{equation}
so that $\delta_\text{nfact}$ is given by
\begin{multline}
\delta_\text{nfact} = - \frac{\alpha}{\pi} \biggl\{ 2 \left[ 1 -
Q_\Pu \ln \left( - \frac{\MW^2}{\hat u} \right) + Q_\Pd \ln \left(
- \frac{\MW^2}{\hat t} \right) \right]
\ln \biggl( \frac{\overline{M}_\PW^2-s}{m_\gamma M_\mathrm{W}}
\biggr) \\
{} - Q_\Pu \Li2 \left( 1 + \frac{\MW^2}{\hat u} \right) +
Q_\Pd \Li2 \left( 1 + \frac{\MW^2}{\hat t} \right) - 2 \biggr\} \text{,}
\end{multline}
where we have made the difference between the Mandelstam variables and their
on-shell-projected counterparts $\hat t, \hat u$ explicit.
For a single resonance with only massless external particles, an
appropriate on-shell projection can be simply realized by the rescaling
\begin{equation}
s \to \frac{\MW^2}{s}\,s = \MW^2, \quad
t \to \frac{\MW^2}{s}\,t = \hat t, \quad
u \to \frac{\MW^2}{s}\,u = \hat u.
\label{eq:DY-OS}
\end{equation}
This result for $\delta_\text{nfact}$ agrees with
the one given for the case of $\mathrm{W}^+$ production in Eq.~(2.22) of
Ref.~\cite{Dittmaier:2001ay}.

For \PZ-boson production the charges are given by
\begin{equation}
Q_1 = Q_2 = Q_q, \quad
Q_3 = Q_4 = Q_\ell = -1, \quad
\overline{Q}_1 = 0.
\end{equation}
The resonance is neutral, so that the contributions $\mathit{mm}$,
$\mathit{mf}$, and $\mathit{im}$ vanish, and the result can be written as
\begin{equation}
\delta_\text{nfact} = -\sum_{a=3,4} \sum_{b=1,2}
\sigma_a Q_a \sigma_b Q_b \,
\frac{\alpha}{\pi} \left\{ 2 \ln \left( \frac{M_\mathrm{Z}^2}{- \hat s_{ab}} \right) \ln \left( \frac{\overline{M}_\PZ^2-s}{m_\gamma
M_\mathrm{Z}} \right) + \Li2 \left( 1 + \frac{M_\mathrm{Z}^2}{\hat s_{ab}} \right) \right\} \text{,}
\end{equation}
where again $\hat s_{ab}$ results from $s_{ab}$ by the on-shell projection~\eqref{eq:DY-OS}
in accordance with Eq.~(2.9) of Ref.~\cite{Dittmaier:2014qza}.

\subsubsection{W-pair production in lepton/hadron/photon collisions}
\label{sec:w-pair-production-in-electron-positron-collisions}

For the case of $\overline{f}_1 f_2 \rightarrow \PWp \PWm \rightarrow f_3
\overline{f}_4 f_5 \overline{f}_6$, we choose the index sets appearing in
Eq.~\eqref{eq:non-factorizable-contribution} to be
\begin{equation}
I = \{ 1,2 \}, \quad R_1 = \{ 3,4 \}, \quad R_2 = \{ 5,6 \}, \quad N = \emptyset
\text{.} \label{eq:w-pair-production-index-sets}
\end{equation}
The corresponding sign factors are
\begin{equation}
\sigma_1 = -1 \text{,} \quad \sigma_2 = 1 \text{,} \quad \sigma_3 = -1 \text{,}
\quad \sigma_4 = 1 \text{,} \quad \sigma_5 = -1 \text{,} \quad \sigma_6 = 1
\end{equation}
and therefore $r=2$ and $n=0$. $N = \emptyset$ means there are no additional
non-resonant particles, so that the sum over $b$ in
Eq.~\eqref{eq:non-factorizable-contribution} simply runs over the initial-state
particles. Furthermore, since $\overline{f}_1$ is the antiparticle of $f_2$, we
have $\sum_{b \in I} \sigma_b Q_b = 0$, so that the contributions from
$\Delta'_\mathrm{mm}$ and $\Delta'_\mathrm{mf}$ cancel
in $\Delta'$ given in Eq.~\eqref{eq:Deltaprime}, because they do not
depend on $b$.

The initial-state contributions, 
i.e.\ the function $\Delta' (i,a;b) =
\Delta_\mathrm{xm} (i,a;b) + \Delta_\mathrm{xf} (i,a;b)$, can be brought into a
form that can be summed together with the remaining non-vanishing contributions,
$\Delta$. To this end, we first define the relative charge of the initial-state
fermions $Q_f = Q_1 = Q_2$ and express their sign factors $\sigma_{1,2}$ in
terms of the charges of the vector bosons,
\begin{equation}
\sigma_1 = - \sigma_2 = \sum_{c \in R_1} \sigma_c Q_c = - \sum_{c \in R_2}
\sigma_c Q_c \text{,} \label{eq:initial-final-charge-identity}
\end{equation}
and then explicitly perform the summation over $i$ and $b$, i.e.\
\begin{equation}
\sum_{i=1}^2 \sum_{a \in R_i} \sum_{b \in I} \sigma_a Q_a \sigma_b Q_b 
\mathrm{Re}\{\Delta' (i,a;b) \}
= \sum_{a \in R_1} \sum_{c \in R_2} \sigma_a Q_a \sigma_c Q_c (-Q_f)
\mathrm{Re}\left\{ \Delta_\mathrm{im}' (a;c) + \Delta_\mathrm{if}' (a;c) \right\} \text{.}
\label{eq:ww-initial-state-contributions}
\end{equation}
On the r.h.s.\ of Eq.~\eqref{eq:ww-initial-state-contributions} we defined two
new functions, one of them
\begin{equation}
\begin{alignedat}{2}
\Delta_\mathrm{im}' (a;c) =
{} & {} \phantom{{}+{}}\Delta_\mathrm{im} (i=1,a;b=1) &
   &  - \Delta_\mathrm{im} (i=1,a;b=2) \\
   &  {} - \Delta_\mathrm{im} (i=2,c;b=1) &
   &  + \Delta_\mathrm{im} (i=2,c;b=2) \text{,}
\end{alignedat}
\end{equation}
in which the summation over $i$ and $b$ is explicit. The definition of
$\Delta_\mathrm{if}'$ is analogous.

As previously constructed, the remaining contributions $\Delta$ have the same
summation structure as the r.h.s.\ of
Eq.~\eqref{eq:ww-initial-state-contributions}, because $i=1$ and $j=2$, so that
with the identity $\sigma_a \sigma_b = (-1)^{a+b}$
Eq.~\eqref{eq:non-factorizable-contribution} reads
\begin{equation}
\delta_\text{nfact} = \sum_{a \in R_1} \sum_{b \in R_2} (-1)^{a+b+1} Q_a Q_b
\frac{\alpha}{\pi} \Re \left\{ \Delta'' \right\} \text{,}
\end{equation}
where we collected all contributions in
\begin{equation}
\Delta'' = \Delta_\mathrm{mm} + \Delta_\mathrm{mf} + \Delta_\mathrm{mm'} +
\Delta_\mathrm{mf'} + \Delta_\mathrm{ff'} - Q_f \left( \Delta_\mathrm{if}' +
\Delta_\mathrm{im}' \right) \text{.}
\label{eq:final-formula-four-fermion-production}
\end{equation}

An on-shell projection is given in Sec.~\ref{sec:on-shell-projection}.  Here we
specialize to the case of two W~bosons and give the on-shell-projected momenta
$\hat{k}_i$ in the centre-of-mass frame. The initial-state momenta are
unmodified,
\begin{equation}
\hat{k}_1 = k_1 \text{,} \quad \hat{k}_2 = k_2 \text{,}
\end{equation}
implying $\overline{\overline{s}}_{12} = s$. Since $M_1 = M_2 = M_\PW$ the
triangle function is $\lambda = s^2 \beta^2$ with the velocity $\beta = \sqrt{1
- 4 M_\PW^2 / s}$. Using the momenta given in Eq.~\eqref{eq:on-shell-resonances}
and fixing the direction $\vec{e}_1 = \vec{\overline{k}}_1 / |
\vec{\overline{k}}_1 |$ of the \PWp~boson, leads to
\begin{equation}
\hat{\overline{k}}_1 = \frac{1}{2} \sqrt{s} \left( 1, \beta \vec{e}_1 \right),
\qquad \hat{\overline{k}}_2 = - \hat{k}_1 - \hat{k}_2 -
\hat{\overline{k}}_1 \text{.}
\label{eq:w-pair-on-shell-projection}%
\end{equation}
Using Eq.~\eqref{eq:on-shell-final-states} and making use of the fact that the
fermions are massless, the on-shell projection reduces to a scale factor for the
momentum whose direction we want to preserve. Since scaling 
massless momenta commutes with
boosts, the scale factor is an invariant and can be easily computed in the
centre-of-mass frame of the vector boson,
\begin{subequations}
\begin{alignat}{2}
\hat{k}_3 &= k_3 \,\frac{M_\mathrm{W}^2}{2 \hat{\overline{k}}_1 k_3}, & \qquad
\hat{k}_4 &= \hat{\overline{k}}_1 - \hat{k}_3 \text{,} \\
\hat{k}_5 &= k_5 \,\frac{M_\mathrm{W}^2}{2 \hat{\overline{k}}_2 k_5}, & \qquad
\hat{k}_6 &= \hat{\overline{k}}_2 - \hat{k}_6 \text{,}
\end{alignat}
\end{subequations}
where the scale factors are derived from the conditions $\hat{k}_4^2 =
\hat{k}_6^2 = 0$. These momenta must be inserted into
Eqs.~\eqref{eq:non-manifestly-nf-contributions}
and~\eqref{eq:manifestly-nf-contributions} which simplifies some of the
kinematical prefactors, e.g.\ $\overline{\overline{s}}_{1 2} \to s$ and
$\tilde{s}_{ia} \to 0$, for all $a \in R_i$. These results agree for the case
$\overline{f}_1 = \Pep, f_2 = \Pem, Q_f = -1$ with the one given in
Refs.~\cite{Denner:1997ia,Denner:2000bj} and for the case of initial-state
quarks with Refs.~\cite{Accomando:2004de,Billoni:2013aba}.

As already mentioned in Sec.~\ref{sec:on-shell-projection}, the chosen on-shell
projection constitutes an intrinsic ambiguity on the method. To determine the
error introduced by this ambiguity and to verify 
that the choice is suitable 
results obtained with different on-shell projections can be compared. 
For $\Pep\Pem\to\PW\PW\to4\,$fermions this check was carried out
in Ref.~\cite{Denner:2000bj}, i.e.\ the results from the
on-shell projection as defined above was
compared against results where the direction of $k_4$ instead $k_3$ was preserved.
The comparison revealed differences from changing the on-shell projections
that are of the order of all other intrinsic uncertainties of the 
double-pole approximation~(DPA), as expected.

For the case of two initial photons, $\gamma \gamma \rightarrow \PWp \PWm$,
there are no initial-state contributions, i.e.\ $Q_f = Q_\gamma = 0$ in
Eq.~\eqref{eq:final-formula-four-fermion-production}. Electroweak corrections to
this process in DPA, 
including these non-factorizable
corrections, were calculated in Ref.~\cite{Bredenstein:2005zk}.

\subsubsection{Vector-boson scattering at hadron colliders}
\label{sec:vector-boson-scattering-at-hadron-colliders}

A prominent process featuring the production of additional non-resonant
particles that were absent in the two previous examples is the case of
vector-boson scattering at hadron colliders.  The production of two vector
bosons that are able to scatter off each other is only possible via radiation
off a quark or an antiquark line, which then subsequently form jets in the final
state. We thus have, at the parton level, the processes $\Pq_1 \Pq_2 \to V V'
\Pq_7 \Pq_8 \to \ell_3 \overline{\ell}_4 \ell_5 \overline{\ell}_6 \Pq_7 \Pq_8$
and all possible combinations with antiquarks that are consistent with charge
conservation. The index sets, thus, are
\begin{equation}
I = \{ 1, 2 \} \text{,} \quad R_1 = \{ 3, 4 \} \text{,} \quad R_2 = \{ 5, 6 \}
\text{,} \quad N = \{ 7, 8 \} \text{.}
\end{equation}
A particularly interesting process is the scattering of same-sign W bosons,
because, e.g., the appearance of $\mu^\pm \mu^\pm$ pairs in an event is a rather
clean event signature and the QCD-initiated production can be efficiently
suppressed by cuts~\cite{Jager:2011ms}.

Although desirable to reach a precision of some percent, electroweak corrections
to this process are are not yet available at this time, due to the fact that the
full correction to a $2 \to 6$ process is extremely challenging. However, as we
argue here, the full correction is also not necessary,
because an evaluation of the corrections in DPA
will certainly be good enough.
In DPA, the vector-boson scattering is a $2 \to 4$ particle production 
process with two resonances followed by two vector-boson decays, so that
the virtual factorizable corrections can be calculated with modern
automated tools for one-loop amplitudes.
The non-factorizable corrections can be
evaluated using our master formula presented in Sec.~\ref{sec:analytic-results}
in a similar fashion as in the examples discussed in the previous sections.

The on-shell projection can be performed as given in
Sec.~\ref{sec:on-shell-projection}. We then keep the momenta of the
initial-state particles and also the momenta of the non-resonant final states,
i.e.\
\begin{equation}
\hat{k}_1 = k_1 \text{,} \quad \hat{k}_2 = k_2 \text{,} \quad \hat{k}_7 = k_7
\text{,} \quad \hat{k}_8 = k_8 \text{.}
\end{equation}
In the centre-of-mass frame of the vector bosons, i.e. the frame where
$\vec{k}_1 + \vec{k}_2 + \vec{k}_7 + \vec{k}_8 = 0$, we can easily construct the
momenta. If the vector bosons have the same mass $M_V$, the momenta are given by
Eq.~\eqref{eq:w-pair-on-shell-projection} if we make the replacements $s \to
\overline{\overline{s}}_{12}$ and $M_\PW \to M_V$. If they have different
masses, e.g.\ in the case of $\PWpm \PZ$ scattering, we can use the general
procedure as given in Eq.~\eqref{eq:on-shell-final-states}.

\section{Conclusion}
\label{sec:conclusion}

Many interesting particle processes at present and potential future high-energy
colliders share the pattern of producing several unstable particles in
intermediate resonant states which decay subsequently, thereby producing
final states of high multiplicities.
At run~2 of the LHC, which has started in 2015, multiple-vector-boson production
such as $\Pp \Pp \to \PW \PW \PW \to \text{6 leptons}$
and massive vector-boson scattering
such as $\Pp \Pp \to \PW \PW + \text{2 jets} \to \text{4 leptons} + \text{2 jets}$
are prominent examples for
corresponding upcoming analyses in the electroweak sector.
In spite of the smaller cross sections of high-multiplicity processes,
predictions for those processes nevertheless have to include
radiative corrections of the strong and electroweak interactions at
next-to-leading order, in order to reach a precision of about $10\%$, or better
since both types of corrections are generically of this size or even larger
in the TeV range.

Calculating radiative corrections to resonance processes poses additional
complications on top of the usual complexity of higher-order calculations,
since gauge invariance is jeopardized by the necessary
Dyson summation of the resonance propagators. For low and intermediate
multiplicities, complete next-to-leading-order calculations are
feasible within the complex-mass scheme, but unnecessarily complicated
and also not needed in view of precision for high multiplicities.
In those cases, predictions where matrix elements are based on
expansions about resonance poles are adequate. Such expansions can be
based on scattering amplitudes directly or on specifically designed
effective field theories. If only the leading contribution of the
expansion is kept, the approach---known as pole approximation---is
particularly intuitive. At next-to-leading order, corrections are
classified into
separately gauge-invariant factorizable and non-factorizable corrections, where
the former can be attributed to the production and decay of the unstable
particles on their mass shell. The remaining non-factorizable corrections are
induced by the exchange of soft photons or gluons
between different production and decay subprocesses.

In this paper, we have presented
explicit analytical results for the non-factorizable
photonic virtual corrections to processes involving an arbitrary number of
unstable particles at the one-loop level. The results represent an essential
building block in the calculation of next-to-leading-order electroweak
corrections in pole approximation and
are ready for a direct implementation in computer codes.
As illustrating examples, we have rederived known results for the single and
pair production of electroweak gauge bosons and have outlined the approach 
for vector-boson scattering.

A generalization of the results to QCD corrections is straightforward
and merely requires the inclusion of the colour flow in the algebraic parts
of the individual contributions, while the analytic part containing the loop
integrals remains the same.

The presented results on virtual non-factorizable corrections help
to close a gap in the ongoing effort of several groups towards
the fully automated calculation of next-to-leading-order corrections,
since the automation of the remaining virtual factorizable
corrections is well under control within QCD with up to 4--6
final-state particles and becomes more and more mature for electroweak
corrections as well.
The situation in view of real QCD and real photonic electroweak corrections
is even better, since tree-level calculations with up to about 10
final-state particles based on
full matrix elements are possible with modern multi-purpose Monte Carlo
generators.
Having at hand generic results on virtual non-factorizable corrections, thus,
opens the door to the fully automated calculation of
virtual corrections to resonance processes in pole approximation.

\section*{Acknowledgements}

We thank Ansgar Denner for carefully reading the manuscript, and
C.S.\ is grateful to Christian Schwinn for useful discussions on
non-factorizable corrections.
This work is supported by the German Science Foundation (DFG)
under contract Di~784/3-1 and via the
Research Training Group GRK~2044.

\begin{appendix}
\section*{Appendix}

\section{Soft-photon emission off resonances}
\label{app:soft-photon}

In this appendix we derive Eqs.~\eqref{eq:Rphoton1} and
\eqref{eq:Rphoton2}, which describe soft-photon emission
off a resonating particle, for particles of spin 0, 1/2, or 1.
Obviously it is sufficient to prove only the first of these equations; the second follows
from the first upon replacing the momentum $\overline{k}_i\to\overline{k}_i-q$,
taking into account that the photon momentum~$q$ is negligible in the numerator.

If the radiating particle $i$ has spin~0, the proof is extremely simple.
Inserting the Feynman rule for the coupling of a scalar particle $i$ to a photon
and for the two scalar propagators, the subdiagram on the l.h.s.\
of Eq.~\eqref{eq:Rphoton1}
can be directly brought to the desired form,
\begin{equation}
\vcenter{\hbox{\includegraphics{feynman-diagrams/current1}}}
\;\;=\;\; \frac{\ri}{K_i(q)}\,(-\ri e \overline{Q}_i) \left(2\overline{k}_i+q\right)^\mu\,\frac{\ri}{K_i}
\;\;\sim\;\;
\frac{2e \overline{Q}_i \overline{k}_i^\mu}{K_i(q)} \;\times\;
\vcenter{\hbox{\includegraphics{feynman-diagrams/propagator1}}} ,
\label{eq:scalarrad}
\end{equation}
where $\sim$ means that the two sides later produce the same soft singularity
structure for small photon momentum~$q$ when embedded in a full diagram. 
According to the arguments of
Sec.~\ref{sec:relevant-feynman-diagrams}, the calculated loop diagram changes by this approximation only
in terms that are not enhanced by resonance $i$.
In Eq.~\eqref{eq:scalarrad} the necessary approximation was just to omit $q$ in the numerator.

If $i$ is a spin-1/2 fermion, inserting the relevant Feynman rules produces
\begin{equation}
\vcenter{\hbox{\includegraphics{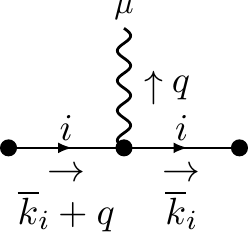}}}
\;\;=\;\; \frac{\ri\left(\dsl{\overline{k}}+\overline{M}_i\right)}{K_i}\,
(-\ri e \overline{Q}_i)\gamma^\mu\,\frac{\ri\left(\dsl{\overline{k}}+\dsl{q}+\overline{M}_i\right)}{K_i(q)}.
\end{equation}
A simple rearrangement of the Dirac matrices leads to the desired form after dropping
again irrelevant (non-resonant) terms,
\begin{align}
\vcenter{\hbox{\includegraphics{feynman-diagrams/Fcurrent}}}
& \;\;\sim\;\;
\ri e \overline{Q}_i\,\frac{\left(\dsl{\overline{k}}_i+\overline{M}_i\right)
\gamma^\mu\left(\dsl{\overline{k}}_i+\overline{M}_i\right)}{K_i(q)K_i}
\;\;=\;\;
\ri e \overline{Q}_i\,\frac{2\overline{k}_i^\mu\left(\dsl{\overline{k}}_i+\overline{M}_i\right)-K_i}{K_i(q)K_i}
\nonumber\\
& \;\;\sim\;\;
\frac{2e \overline{Q}_i \overline{k}_i^\mu}{K_i(q)} \;\times\;
\vcenter{\hbox{\includegraphics{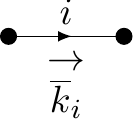}}} .
\end{align}
Analogous manipulations with the opposite fermion flow for an antifermion resonance
produce the same result.

The case where $i$ is a charged spin-1 boson deserves more care.
We assume that $i$ is a gauge boson that receives its mass by the Higgs mechanism,
just like the W~boson in the SM. In principle, we, thus, have to consider
all possible loop diagrams with subdiagrams~\eqref{eq:Rphoton1}, in which the resonance
line $i$ represents the gauge-boson field,
its corresponding would-be Goldstone boson, or even a Faddeev--Popov ghost
field. However, if we switch to an $R_\xi$~gauge for the $i$ field where its
gauge parameter $\xi_i\ne1$, the propagators of the corresponding Goldstone and ghost fields
develop their pole at $\overline{k}_i^2=\xi_i\overline{M}_i^2\ne \overline{M}_i^2$.
However, a pole at $\overline{k}_i^2=\overline{M}_i^2$ would be necessary to
produce soft divergences on resonance which in turn is a necessary condition
for the corresponding diagrams to contribute to non-factorizable corrections.
Consequently, we can ignore subgraphs~\eqref{eq:Rphoton1} with would-be
Goldstone boson or ghost fields in the following.
In the adopted $R_\xi$~gauge the $i$~propagator is given by
\begin{equation}
\vcenter{\hbox{\includegraphics{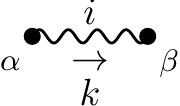}}}
\;\;=\;\;
G_{i,\alpha\beta}(k)
\;=\;
\frac{-\ri\left(g_{\alpha\beta}-\frac{k_\alpha k_\beta}{k^2}\right)}{k^2-\overline{M}_i^2}
+\frac{-\ri\xi_i\frac{k_\alpha k_\beta}{k^2}}{k^2-\xi_i\overline{M}_i^2}.
\end{equation}
Obviously the second term with the unphysical pole at
$\overline{k}_i^2=\xi_i\overline{M}_i^2$ again does not contribute
to the non-factorizable corrections and can be ignored.
Inserting the respective Feynman rules, we obtain
\begin{align}
\vcenter{\hbox{\includegraphics{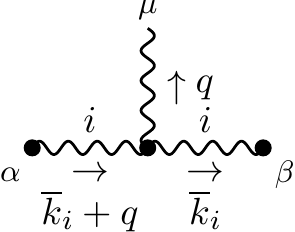}}}
\;\;=\;\; &
(-\ri \overline{Q}_i e) \left[ g^{\mu\nu}\left(\overline{k}_i+2q\right)^\rho
+ g^{\nu\rho}\left(-2\overline{k}_i-q\right)^\mu
+ g^{\rho\mu}\left(\overline{k}_i-q\right)^\nu
\right]
\nonumber\\[-2em]
& \times
G_{i,\alpha\nu}\left(\overline{k}_i+q\right)
G_{i,\rho\beta}\left(\overline{k}_i\right)
\nonumber\\
\;\;\sim\;\; &
\ri \overline{Q}_i e \left[ g^{\mu\nu}\overline{k}_i^\rho
-2g^{\nu\rho}\overline{k}_i^\mu
+g^{\rho\mu}\overline{k}_i^\nu
\right] \,
\frac{g_{\alpha\nu}-\frac{\overline{k}_{i,\alpha}\overline{k}_{i,\nu}}{\overline{k}_i^2}}{K_i(q)} \;
\frac{g_{\rho\beta}-\frac{\overline{k}_{i,\rho}\overline{k}_{i,\beta}}{\overline{k}_i^2}}{K_i}
\nonumber\\
\;\;=\;\; &
-2\ri \overline{Q}_i e \overline{k}_i^\mu \;
\frac{g_{\alpha\beta}-\frac{\overline{k}_{i,\alpha}\overline{k}_{i,\beta}}{\overline{k}_i^2}}{K_i(q)K_i}
\nonumber\\
\;\;\sim\;\; &
\frac{2e \overline{Q}_i \overline{k}_i^\mu}{K_i(q)} \;\times\;
\vcenter{\hbox{\includegraphics{feynman-diagrams/Vpropagator}}} \; ,
\end{align}
where we have neglected $q$ in the numerator in the first $\sim$ relation and
performed simple four-vector contractions in the subsequent step.
The final $\sim$ relation, which proves Eq.~\eqref{eq:Rphoton1},
is again valid up to irrelevant terms with an unphysical propagator pole.

\section{Derivation of virtual non-factorizable corrections}
\label{app:virtual-corrections}

In this appendix we calculate the non-factorizable corrections induced
by the various diagram types shown in Fig.~\ref{fig:non-factorizable-diagrams},
making use of the generic results derived in
Sec.~\ref{sec:derivation-of-the-non-factorizable-corrections}, which
are summarized in Eq.~\eqref{eq:deltanfact}.
Our aim is to express all contributions in terms of known
standard scalar one-loop integrals as defined in App.~\ref{app:scalar-integrals}.
\\[1em]
List of the different types of non-factorizable corrections:

\begin{itemize}
\item The $\mathit{ff'}$-diagram in Fig.~\ref{fig:ffprime} is manifestly
non-factorizable and involves the following combination of currents
\begin{equation}
j_{a}(q) \cdot j_{b}(-q) = \frac{2e \sigma_a Q_a k_a^\mu}{q^2+2qk_a} \,
\frac{K_i}{K_i(q)} \frac{2e \sigma_b Q_b k_{b,\mu}}{q^2-2qk_b} \,
\frac{K_j}{K_j(-q)},
\end{equation}
where $a \in R_i$ and $b \in R_j$ are decay particles of two different
resonances $i,j\in \overline{R}$, $i \neq j$.  Inserting this into the
integral~\eqref{eq:deltanfact} and using $e^2=4\pi\alpha$,
directly leads to the contribution
$\delta_\mathrm{ff'} (i,a;j,b)= - \frac{\alpha}{\pi} \sigma_a Q_a \sigma_b Q_b
\Re \left\{ \Delta_\mathrm{ff'} (i,a;j,b) \right\}$ with
\begin{equation}
\begin{split}
\Delta_\mathrm{ff'} (i,a;j,b) \,=\, &
- (s_{ab}-m_a^2-m_b^2) \, K_i K_j \\
& {} \times E_0(k_a,\overline{k}_i,-\overline{k}_j,-k_b,m_\gamma^2,
m_a^2,\overline{M}_i^2,\overline{M}_j^2,m_b^2).
\end{split}
\end{equation}
The sum over all non-equivalent pairs $i,j$ and corresponding pairs $a,b$ is
\begin{equation}
\delta_\mathrm{ff'} = - \frac{\alpha}{\pi} \sum_{i=1}^r \sum_{j=i+1}^r \sum_{a
\in R_i} \sum_{b \in R_j} \sigma_a Q_a \sigma_b Q_b \Re \left\{
\Delta_\mathrm{ff'} (i,a;j,b) \right\} \text{.}
\end{equation}

\item The $\mathit{xf}$-diagrams in Fig.~\ref{fig:nf} and Fig.~\ref{fig:if} are
manifestly non-factorizable and involve the following combination of currents,
\begin{equation}
j_{a}(q) \cdot j_{b}(-q) = \frac{2e \sigma_a Q_a k_a^\mu}{q^2+2qk_a} \,
\frac{K_i}{K_i(q)} \frac{2e \sigma_b Q_b k_{b,\mu}}{q^2-2qk_b},
\end{equation}
where $a \in R_i$, $i\in \overline{R}$ and $b \in I\cup N$.  Inserting this into
the integral~\eqref{eq:deltanfact}, directly leads to the contribution
$\delta_\mathrm{xf} (i,a;b)= - \frac{\alpha}{\pi} \sigma_a Q_a \sigma_b Q_b \Re
\left\{ \Delta_\mathrm{xf} (i,a;b) \right\}$ with
\begin{equation}
\begin{split}
\Delta_\mathrm{xf} (i,a;b) &=
- (s_{ab}-m_a^2-m_b^2) \, K_i \, 
D_0(k_a,\overline{k}_i,-k_b,m_\gamma^2, m_a^2,\overline{M}_i^2,m_b^2) \\
&= - (s_{ab}-m_a^2-m_b^2) \, K_i \, 
D_0( m_a^2, \widetilde{s}_{ia}, \overline{s}_{ib}, m_b^2,
\overline{k}_i^2, s_{ab},
m_\gamma^2, m_a^2,\overline{M}_i^2,m_b^2).
\end{split}
\end{equation}
The sum over all resonances~$i$ and corresponding pairs $a,b$ is
\begin{equation}
\delta_\mathrm{xf} = - \frac{\alpha}{\pi} \sum_{i=1}^r \sum_{a \in R_i} \sum_{b
\in I\cup N} \sigma_a Q_a \sigma_b Q_b \Re \left\{ \Delta_\mathrm{xf} (i,a;b)
\right\} \text{.}
\end{equation}

\item The $\mathit{mf'}$-diagram in Fig.~\ref{fig:mfprime} is manifestly
non-factorizable and receives contributions from the following combination of
currents,
\begin{equation}
\begin{split}
\overline{j}_{\inc,i}(q) \cdot j_{b}(-q) + \overline{j}_{\out,i}(q) \cdot
j_{b}(-q) \,=\, & \frac{2e \overline{Q}_i
\overline{k}_i^\mu}{q^2+2\overline{k}_i q} \left(\frac{K_i}{K_i(q)}-1\right)
\frac{2e \sigma_b Q_b k_{b,\mu}}{q^2-2qk_b} \, \frac{K_j}{K_j(-q)} \\
\,=\, &
- \frac{4e^2 \overline{Q}_i \sigma_b Q_b \, \overline{k}_i\cdot k_b \, K_j}
  {K_i(q) (q^2-2qk_b)K_j(-q)},
\end{split}
\end{equation}
where $i,j\in \overline{R}$ are different resonances ($i\ne j$) and $b \in R_j$.
Inserting this into the integral~\eqref{eq:deltanfact}, leads to the
contribution $\delta_\mathrm{mf'} (i;j,b)= \frac{\alpha}{\pi} \overline{Q}_i
\sigma_b Q_b \Re \left\{ \Delta'_\mathrm{mf'} (i;j,b) \right\}$ with
\begin{equation}
\begin{split}
\Delta'_\mathrm{mf'} (i;j,b) &=
- (\overline{s}_{ib}-\overline{k}_i^2-m_b^2) \, K_j\,
  D_0(\overline{k}_i,-\overline{k}_j,-k_b,m_\gamma^2,
\overline{M}_i^2,\overline{M}_j^2,m_b^2) \\
& \sim
- (\overline{s}_{ib}-M_i^2-m_b^2) \, K_j\,
  D_0(\overline{k}_i,-\overline{k}_j,-k_b,m_\gamma^2,
\overline{M}_i^2,\overline{M}_j^2,m_b^2) \\
& =
- (\overline{s}_{ib}-M_i^2-m_b^2) \, K_j\,
  D_0(\overline{k}_i^2, \overline{\overline{s}}_{ij},
\widetilde{s}_{jb}, m_b^2, \overline{k}_j^2, \overline{s}_{ib},
m_\gamma^2,\overline{M}_i^2,\overline{M}_j^2,m_b^2).
\end{split}
\end{equation}
The sum over the resonances $i$ and $j$ and its decay product $b$ is
\begin{equation}
\begin{split}
\delta_\mathrm{mf'} &= \frac{\alpha}{\pi} \sum_{i=1}^r
\sum_{\substack{j=1\\j \neq i}}^r \sum_{b
\in R_j} \overline{Q}_i \sigma_b Q_b \Re \left\{ \Delta'_\mathrm{mf'} (i;j,b)
\right\} \\
&= - \frac{\alpha}{\pi} \sum_{i=1}^r \sum_{j=i+1}^r \sum_{a \in R_i} \sum_{b \in
R_j} \sigma_a Q_a \sigma_b Q_b \bigl( \Re \left\{ \Delta'_{\mathrm{mf}'} (i;j,b)
\right\} + \Re \left\{ \Delta'_{\mathrm{mf}'} (j;i,a) \right\} \bigr) \text{.}
\end{split}
\end{equation}

\item The $\mathit{mf}$-diagram in Fig.~\ref{fig:mf} is not manifestly
non-factorizable, since it contains a factorizable part. The non-factorizable
part receives contributions from the following combination of currents,
\begin{equation}
\begin{split}
\widetilde{j}_{\out,i}(q) \cdot j_{a}(-q) \,=\, & \frac{2e \overline{Q}_i
\overline{k}_i^\mu}{q^2-2\overline{k}_i q} \, \frac{2e \sigma_a Q_a
k_{a,\mu}}{q^2-2qk_a}\, \frac{K_i}{K_i(-q)} \\
\,=\, & \frac{4e^2 \overline{Q}_i\sigma_a Q_a  \,\overline{k}_i \cdot
k_a}{q^2-2qk_a} \left( \frac{1}{q^2-2\overline{k}_i q} - \frac{1}{K_i(-q)}
\right),
\end{split}
\end{equation}
where $i\in \overline{R}$ and $a \in R_i$.  Inserting this into the
integral~\eqref{eq:deltanfact}, leads to the contribution $\delta_\mathrm{mf}
(i;a)= - \frac{\alpha}{\pi} \overline{Q}_i \sigma_a Q_a \Re \left\{
\Delta_\mathrm{mf}' (i;a) \right\}$ with
\begin{equation}
\begin{split}
\Delta_\mathrm{mf}' (i;a) & = (\tilde{s}_{ia}-\overline{k}_i^2-m_a^2) \, \Bigl\{
C_0(-\overline{k}_i,-k_a,m_\gamma^2, \overline{k}_i^2,m_a^2) \\
& \hspace*{11em} {} - C_0(-\overline{k}_i,-k_a,m_\gamma^2,
\overline{M}_i^2,m_a^2) \Bigr\} \\
& \sim (\tilde{s}_{ia}-M_i^2-m_a^2) \, \Bigl\{
C_0(M_i^2,\tilde{s}_{ia},m_a^2,m_\gamma^2, M_i^2,m_a^2) \\
& \hspace*{11em} {} - C_0(\overline{k}_i^2,\tilde{s}_{ia},m_a^2,0,
\overline{M}_i^2,m_a^2) \Bigr\},
\end{split}
\end{equation}
with the usual difference between the full off-shell diagram and its
factorizable part with $\overline{k}_i^2=M_i^2$.
The final form, where invariants are used as arguments of the $C_0$ integrals,
makes the appearance of off-shell and on-shell momenta on the resonance lines
better visible.
The sum over all resonances $i$
and its decay products $a$ is
\begin{equation}
\begin{split}
\delta_\mathrm{mf} = - &\frac{\alpha}{\pi} \sum_{i=1}^r \sum_{a \in R_i}
\overline{Q}_i \sigma_a Q_a \Re \left\{ \Delta_\mathrm{mf}' (i,a) \right\} \\
= - &\frac{\alpha}{\pi} \sum_{i=1}^r \sum_{a \in R_i} \left(
\sum_{\substack{j=1\\j \neq i}}^r
\sum_{b \in R_j} + \sum_{b \in I \cup N} \right) \sigma_a Q_a \sigma_b Q_b \Re
\left\{ \Delta_\mathrm{mf}' (i,a) \right\} \\
= - &\frac{\alpha}{\pi} \sum_{i=1}^r
\sum_{j=i+1}^r
\sum_{a \in R_i}
\sum_{b \in
R_j} \sigma_a Q_a \sigma_b Q_b \bigl( \Re \left\{ \Delta_\mathrm{mf}' (i,a)
\right\} + \Re \left\{ \Delta_\mathrm{mf}' (j,b) \right\} \bigr) \\
  - &\frac{\alpha}{\pi} \sum_{i=1}^r \sum_{a \in R_i} \sum_{b \in I \cup N}
\sigma_a Q_a \sigma_b Q_b \Re \left\{ \Delta_\mathrm{mf}' (i,a) \right\}
\text{.}
\end{split}
\end{equation}

\item The $\mathit{xm}$-diagrams in Fig.~\ref{fig:mn} and Fig.~\ref{fig:im} are
not manifestly non-factorizable, since they contain factorizable contributions
as well. The non-factorizable part receives contributions from the following
combination of currents,
\begin{equation}
\begin{split}
\overline{j}_{\inc,i}(q) \cdot j_{b}(-q) \,=\, & \frac{2e \overline{Q}_i
\overline{k}_i^\mu}{q^2+2\overline{k}_i q} \, \frac{K_i}{K_i(q)} \frac{2e
\sigma_b Q_b k_{b,\mu}}{q^2-2qk_b} \\
\,=\, & \frac{4e^2 \overline{Q}_i\sigma_b Q_b  \,\overline{k}_i \cdot
k_b}{q^2-2qk_b} \left( \frac{1}{q^2+2\overline{k}_i q} - \frac{1}{K_i(q)}
\right),
\end{split}
\end{equation}
where $i\in \overline{R}$ and $b \in I\cup N$.  Inserting this into the
integral~\eqref{eq:deltanfact}, leads to the contribution $\delta_\mathrm{xm}
(i;b)= \frac{\alpha}{\pi} \overline{Q}_i \sigma_b Q_b \Re \left\{
\Delta_\mathrm{xm} (i;b) \right\}$ with
\begin{equation}
\begin{split}
\Delta_\mathrm{xm} (i;b) & = (\overline{s}_{ib}-\overline{k}_i^2-m_b^2) \,
\Bigl\{ C_0(\overline{k}_i,-k_b,m_\gamma^2, \overline{k}_i^2,m_b^2) \\
& \hspace*{11em} {} - C_0(\overline{k}_i,-k_b,m_\gamma^2,
\overline{M}_i^2,m_b^2) \Bigr\} \\
& \sim (\overline{s}_{ib}-M_i^2-m_b^2) \, \Bigl\{
C_0(M_i^2,\overline{s}_{ib},m_b^2,m_\gamma^2, M_i^2,m_b^2) \\
& \hspace*{11em} {} - C_0(\overline{k}_i^2,\overline{s}_{ib},m_b^2, 0,
\overline{M}_i^2,m_b^2) \Bigr\},
\end{split}
\end{equation}
which again reflects the subtraction of the factorizable part with an on-shell
momentum of the resonance ($\overline{k}_i^2=M_i^2$) from the full off-shell
diagram.  The sum over all resonances $i$ and other particles~$b$ of the
production process reads
\begin{equation}
\begin{split}
\delta_\mathrm{xm} &= \frac{\alpha}{\pi} \sum_{i=1}^r \sum_{b \in I \cup N}
\overline{Q}_i \sigma_b Q_b \Re \left\{ \Delta_\mathrm{xm} (i;b) \right\} \\
&= - \frac{\alpha}{\pi} \sum_{i=1}^r \sum_{a \in R_i} \sum_{b \in I \cup N}
\sigma_a Q_a \sigma_b Q_b \Re \left\{ \Delta_\mathrm{xm} (i;b) \right\} \text{.}
\end{split}
\end{equation}

\item
The $\mathit{mm'}$-diagram in Fig.~\ref{fig:mmprime} is not manifestly
non-factorizable, i.e.\ it contains both factorizable and non-factorizable
parts. Its non-factorizable contribution involves the following combinations of
ESPA currents,
\begin{align}
\lefteqn{
\overline{j}_{\out,i}(q) \cdot \overline{j}_{\inc,j}(-q)
+
\overline{j}_{\inc,i}(q) \cdot \overline{j}_{\out,j}(-q)
+
\overline{j}_{\inc,i}(q) \cdot \overline{j}_{\inc,j}(-q)
} \hspace*{2em}&
\nonumber\\
= \;&
- \frac{2e \overline{Q}_i \overline{k}_i^\mu}{q^2+2\overline{k}_i q}
\frac{2e \overline{Q}_j \overline{k}_{j,\mu}}{q^2-2\overline{k}_j q} \,
\frac{K_j}{K_j(-q)}
-
\frac{2e \overline{Q}_i \overline{k}_i^\mu}{q^2+2\overline{k}_i q} \,
\frac{K_i}{K_i(q)}
\frac{2e \overline{Q}_j \overline{k}_{j,\mu}}{q^2-2\overline{k}_j q}
\nonumber\\
& {} +
\frac{2e \overline{Q}_i \overline{k}_i^\mu}{q^2+2\overline{k}_i q} \,
\frac{K_i}{K_i(q)}
\frac{2e \overline{Q}_j \overline{k}_{j,\mu}}{q^2-2\overline{k}_j q} \,
\frac{K_j}{K_j(-q)}
\nonumber\\
= \;&
4e^2 \overline{Q}_i \overline{Q}_j (\overline{k}_i\cdot \overline{k}_j)
\left(
\frac{1}{K_i(q)} \,
\frac{1}{K_j(-q)} \,
-
\frac{1}{q^2+2\overline{k}_i q} \,
\frac{1}{q^2-2\overline{k}_j q}
\right),
\end{align}
where $i,j\in \overline{R}$ are different resonances, $i\ne j$. We have used
Eqs.~\eqref{eq:partfract1} and \eqref{eq:partfract2} to obtain the final form.
Inserting this into Eq.~\eqref{eq:deltanfact}, we
obtain its contribution $\delta_\mathrm{mm'} (i;j) = - \frac{\alpha}{\pi}
\overline{Q}_i \overline{Q}_j \Re \left\{ \Delta_\mathrm{mm'} (i;j) \right\}$ to
$\delta_\text{nfact}$, where
\begin{equation}
\begin{split}
\Delta_\mathrm{mm'} (i;j) \;=\;& -\left(\overline{\overline{s}}_{ij} - \overline{k}_i^2 - \overline{k}_j^2\right) \,
\\
& {} \times \left\{
 C_0 \left(\overline{k}_i,-\overline{k}_j, m_\gamma^2 ,\overline{M}_i^2,\overline{M}_j^2\right)
-C_0 \left(\overline{k}_i,-\overline{k}_j,m_\gamma^2,\overline{k}_i^2,\overline{k}_j^2\right)
\right\} \\
\;\sim\;&
-
\left(\overline{\overline{s}}_{ij} - M_i^2 - M_j^2\right) \, \\
& {} \times \left\{
 C_0 \left(\overline{k}_i^2,\overline{\overline{s}}_{ij},\overline{k}_j^2,
0,\overline{M}_i^2,\overline{M}_j^2\right)
-C_0 \left(M_i^2,\overline{\overline{s}}_{ij},M_j^2,m_\gamma^2,M_i^2,M_j^2\right)
\right\},
\end{split}
\end{equation}
where $\sim$ again means identical up to non-resonant terms.
The final form nicely shows how the subtraction of the
factorizable part, where the resonance momenta $\overline{k}_{i,j}$ are on shell,
from the full diagram defines the non-factorizable contribution.
Summing over all resonance pairs $i,j$ and
using charge conservation in the form~\eqref{eq:local-charge-conservation},
the full $\mathit{mm'}$ contribution can be written as
\begin{equation}
\begin{split}
\delta_\mathrm{mm'} &= - \frac{\alpha}{\pi} \sum_{i=1}^r \sum_{j=i+1}^r
\overline{Q}_i \overline{Q}_j \Re \left\{ \Delta_\mathrm{mm'} (i;j) \right\} \\
&= - \frac{\alpha}{\pi} \sum_{i=1}^r \sum_{j=i+1}^r \sum_{a \in R_i} \sum_{b \in
R_j} \sigma_a Q_a \sigma_b Q_b \Re \left\{ \Delta_\mathrm{mm'} (i;j) \right\}
\text{.}
\end{split}
\end{equation}

\item
The $\mathit{mm}$-diagram in Fig.~\ref{fig:mm} deserves some particular care,
since it should be considered in combination with its contribution to the mass
renormalization counterterm of resonance~$i$.  According to the ESPA currents,
the following combination of currents defines the non-factorizable contribution,
\begin{equation}
\begin{split}
\widetilde{j}_{\out,i}(q) \cdot \overline{j}_{\inc,i}(-q) &=
\frac{2e \overline{Q}_i \overline{k}_i^\mu}{q^2-2\overline{k}_i q} \,
\frac{2e \overline{Q}_i \overline{k}_{i,\mu}}{q^2-2\overline{k}_i q} \,
\frac{K_i}{K_i(-q)}
\\
&= 4e^2 \overline{Q}_i^2 \overline{k}_i^2
\left(
\frac{1}{K_i K_i(-q)}
-\frac{1}{K_i(q^2-2\overline{k}_i q)}
+\frac{1}{(q^2-2\overline{k}_i q)^2}
\right)
\end{split}
\end{equation}
for all $i\in \overline{R}$.
Inserting this into the integral~\eqref{eq:deltanfact},
leads to the contribution
$\delta_\mathrm{mm} (i)= \frac{\alpha}{\pi} \overline{Q}_i^2 \Re \left\{
\Delta_\mathrm{mm}' (i) \right\}$ 
with%
\footnote{The term $\propto 1/(q^2-2\overline{k}_i q)^2$ can be identified
with the momentum derivative
$B'_0(p_1^2,m_\gamma^2,m_1^2)=\partial B_0(p_1^2,m_\gamma^2,m_1^2)/\partial p_1^2$
by applying
$\partial/\partial p_1^2=1/(2p_1^2)p_1^\mu\partial/\partial p_1^\mu$
as follows:
$\partial[1/(q^2+2p_1 q+p_1^2-m_1^2)]/\partial p_1^2 =
-(qp_1+p_1^2)/(q^2+2p_1 q+p_1^2-m_1^2)^2/p_1^2
\sim
-1/(q^2+2p_1 q+p_1^2-m_1^2)^2$.
}
\begin{equation}
\begin{split}
\Delta_\mathrm{mm}' (i) & =
2\overline{k}_i^2 \,
\left\{
\frac{B_0\left( \overline{k}_i^2,m_\gamma^2,\overline{M}_i^2 \right)
-B_0\left( \overline{k}_i^2,m_\gamma^2,\overline{k}_i^2 \right)}{K_i}
-B'_0\left( \overline{k}_i^2,m_\gamma^2,\overline{k}_i^2 \right)
\right\}
\\
& \sim
2M_i^2 \,
\left\{
\frac{B_0\left( \overline{k}_i^2,0,\overline{M}_i^2 \right)
-B_0\left( \overline{M}_i^2,m_\gamma^2,\overline{M}_i^2 \right)}{K_i}
-B'_0\left( M_i^2,m_\gamma^2,M_i^2 \right)
\right\}.
\end{split}
\end{equation}
This final form can be interpreted in two different ways:
Taking the first $B_0$ term as the full off-shell contribution, the second and third
terms correspond to its on-shell subtraction to obtain its non-factorizable part.
Performing the same subtraction for the corresponding counterterm contribution
connected with the $i$~self-energy, gives zero, because there is no issue
with respect to interchanging limits in the loop integration, since
the renormalization constants are always calculated first.
The alternative interpretation is to consider the terms in the
curly brackets as the full off-shell contribution of the photon-exchange
diagram and the corresponding counterterms, where the last-but-one and the
last terms correspond to the mass and wave-function renormalization of the
$i$~line in the on-shell renormalization scheme, respectively.
By construction, in this scheme on-shell particles do not receive self-energy corrections,
i.e.\ the factorizable part of the considered contribution in
curly brackets is zero, in accordance with our result.

Summation over all resonances~$i$ leads to
\begin{equation}
\begin{split}
\delta_\mathrm{mm} = + &\frac{\alpha}{\pi} \sum_{i=1}^r \overline{Q}_i^2 \Re
\left\{ \Delta_\mathrm{mm}' (i) \right\} = - \frac{\alpha}{\pi} \sum_{i=1}^r
\sum_{a \in R_i} \sigma_a Q_a \overline{Q}_i \Re \left\{ \Delta_\mathrm{mm}' (i)
\right\} \\
= - &\frac{\alpha}{\pi} \sum_{i=1}^r
\sum_{a \in R_i} \left(
\sum_{\substack{j=1\\j \neq i}}^r
\sum_{b \in R_j} + \sum_{b \in I \cup N} \right) \sigma_a Q_a \sigma_b Q_b \Re
\left\{ \Delta_\mathrm{mm}' (i) \right\} \\
= - &\frac{\alpha}{\pi} \sum_{i=1}^r
\sum_{j=i+1}^r  \sum_{a \in R_i} \sum_{b \in
R_j} \sigma_a Q_a \sigma_b Q_b \bigl( \Re \left\{ \Delta_\mathrm{mm}' (i)
\right\} + \Re \left\{ \Delta_\mathrm{mm}' (j) \right\} \bigr) \\
- &\frac{\alpha}{\pi} \sum_{i=1}^r \sum_{a \in R_i} \sum_{b \in I \cup N}
\sigma_a Q_a \sigma_b Q_b \Re \left\{ \Delta_\mathrm{mm}' (i) \right\} \text{.}
\end{split}
\end{equation}

\end{itemize}

\section{Scalar Integrals}
\label{app:scalar-integrals}

The scalar integrals used in this paper are defined as
\begin{subequations}
\begin{align}
B_0 \left( p_1, m_\gamma^2, m_1^2 \right) &= (2\pi\mu)^{4-D}\int \frac{\mathrm{d}^D q}{\mathrm{i}
\pi^2} \frac{1}{q^2 - m_\gamma^2} \frac{1}{(q+p_1)^2 - m_1^2} \text{,} \\
C_0 \left( p_1, p_2, m_\gamma^2, m_1^2, m_2^2 \right) &= (2\pi\mu)^{4-D}\int \frac{\mathrm{d}^D
q}{\mathrm{i} \pi^2} \frac{1}{q^2 - m_\gamma^2} \prod_{i=1}^2 \frac{1}{(q+p_i)^2
- m_i^2} \text{,} \\
D_0 \left( p_1, p_2, p_3, m_\gamma^2, m_1^2, m_2^2, m_3^2 \right) &= (2\pi\mu)^{4-D}\int
\frac{\mathrm{d}^D q}{\mathrm{i} \pi^2} \frac{1}{q^2 - m_\gamma^2} \prod_{i=1}^3
\frac{1}{(q+p_i)^2 - m_i^2} \text{,} \\
E_0 \left( p_1, p_2, p_3, p_4, m_\gamma^2, m_1^2, m_2^2, m_3^2, m_4^2 \right) &= (2\pi\mu)^{4-D}\int
\frac{\mathrm{d}^D q}{\mathrm{i} \pi^2} \frac{1}{q^2 - m_\gamma^2} \prod_{i=1}^4
\frac{1}{(q+p_i)^2 - m_i^2} \text{,}
\end{align}
\end{subequations}
and
\begin{equation}
B_0' \left( p_1, m_\gamma^2, m_1^2 \right) =
\frac{\partial}{\partial p_1^2} B_0 \left( p_1, m_\gamma^2, m_1^2 \right),
\end{equation}
which is used in the ${mm}$ contribution.  
The integrals are defined in $D=4-2\epsilon$ dimensions
in order to regularize the UV divergence in the $B_0$ function and
(if relevant) to regularize possible IR (soft and collinear)
singularities in the other functions.
The scale $\mu$ represents the arbitrary reference scale of dimensional
regularization.
Sometimes it is convenient
to give the arguments of the loop functions in terms of invariants parametrizing
the integral, as e.g.\
\begin{subequations}
\begin{align}
B_0 \left( p_1, m_\gamma^2, m_1^2 \right) &\equiv
B_0 \left( p_1^2, m_\gamma^2, m_1^2 \right)
\text{,} \\
C_0 \left( p_1, p_2, m_\gamma^2, m_1^2, m_2^2 \right) &\equiv
C_0 \left( p_1^2, (p_2-p_1)^2, p_2^2, m_\gamma^2, m_1^2, m_2^2 \right)
\text{,} \\
D_0 \left( p_1, p_2, p_3, m_\gamma^2, m_1^2, m_2^2, m_3^2 \right)
\nonumber\\
& \hspace*{-10em} \equiv D_0 \left( p_1^2, (p_2-p_1)^2, (p_3-p_2)^2, p_3^2, p_2^2, (p_3-p_1)^2,
m_\gamma^2, m_1^2, m_2^2, m_3^2 \right)
\text{.}
\end{align}
\end{subequations}

For the kinematical case considered in 
Eq.~\eqref{eq:non-resonant-initial-final-contribution},
i.e.\ for massless external particles 
($m_a, m_b \to 0$ with $m_a, m_b\gg m_\gamma \to 0$),
the integrals necessary for
Eq.~\eqref{eq:non-resonant-initial-final-contribution} are given in the
following. The relation '$\sim$' implies that we performed the on-shell
projection $\overline{k}_i^2 \to M_i^2$ and set the masses to the real ones,
$\overline{M}_i^2 \to M_i^2$, whenever
possible. In places where the
propagator denominator appears inside a logarithm, $K_i = \overline{k}_i^2 -
\overline{M}_i^2$, this is not possible, and $K_i$ is kept with its full
dependence on the original
momentum $\overline{k}_i^2$. The on-shell
projection of the invariants, e.g.\ $\overline{s}_{ib} \to
\hat{\overline{s}}_{ib}$, is implicitly understood to keep the notation brief.
The relevant integrals explicitly read
\begin{subequations}
\begin{multline}
D_0 \left( m_b^2, \overline{s}_{ib}, \widetilde{s}_{ia}, m_a^2,
\overline{k}_i^2, s_{ab}, m_\gamma^2, m_b^2, \overline{M}_i^2, m_a^2  \right)
\\
\sim \frac{1}{s_{ab} K_i} \left\{ 2 \ln \left( \frac{m_a m_b}{- s_{ab}} \right) \ln
\left( \frac{m_\gamma M_i}{- K_i} \right) - \ln^2 \left( \frac{m_b M_i}{M_i^2 -
\overline{s}_{ib}} \right) - \ln^2 \left( \frac{m_a M_i}{M_i^2 -
\widetilde{s}_{ia}} \right) \right. \\
\left. 
- \cLi2\left( \frac{M_i^2 - \widetilde{s}_{ia}}{M_i^2} ,
\frac{M_i^2 - \overline{s}_{ib}}{-s_{ab}} \right) 
- \frac{\pi^2}{3} 
\right\} \text{,}
\end{multline}
\begin{multline}
C_0 \left( \overline{k}_i^2, \overline{s}_{ib}, m_b^2, 0, \overline{M}_i^2,
m_b^2 \right) - C_0 \left( M_i^2 , \overline{s}_{ib}, m_b^2, m_\gamma^2, M_i^2,
m_b^2 \right) \\
\sim \frac{1}{\overline{s}_{ib} - M_i^2} \left\{ 
\ln \left( \frac{m_b M_i}{M_i^2 - \overline{s}_{ib}} \right) 
\left[ 2\ln \left( \frac{- K_i}{m_\gamma M_i} \right) 
+\ln \left( \frac{m_b M_i}{M_i^2 - \overline{s}_{ib}} \right) \right]
+ \frac{\pi^2}{6} \right\} \text{,}
\end{multline}
\begin{multline}
C_0 \left( \overline{k}_i^2, \widetilde{s}_{ia}, m_a^2, 0, \overline{M}_i^2,
m_a^2 \right) - C_0 \left( M_i^2, \widetilde{s}_{ia}, m_a^2, m_\gamma^2, M_i^2,
m_a^2 \right) \\
\sim \frac{1}{\widetilde{s}_{ia} - M_i^2} \left\{ 
\ln \left( \frac{m_a M_i}{M_i^2 - \widetilde{s}_{ia}} \right) 
\left[ 2\ln \left( \frac{- K_i}{m_\gamma M_i} \right)
+\ln \left( \frac{m_a M_i}{M_i^2 - \widetilde{s}_{ia}} \right) \right]
+ \frac{\pi^2}{6} \right\} \text{,}
\end{multline}
\begin{equation}
\frac{B_0 \left( \overline{k}_i^2, 0, \overline{M}_i^2 \right) - B_0 \left(
\overline{M}_i^2, m_\gamma^2, \overline{M}_i^2 \right)}{K_i} - B'_0 \left(
M_i^2, m_\gamma^2, M_i^2 \right) \sim \frac{1}{M_i^2} \left\{ \ln \left(
\frac{m_\gamma M_i}{- K_i} \right) + 1 \right\} \text{,}
\end{equation}
\end{subequations}
where $s_{ab}$, $\widetilde{s}_{ia}$, and $\overline{s}_{ib}$ are implicitly understood
as $s_{ab}+\ri0$, $\widetilde{s}_{ia}+\ri0$, and $\overline{s}_{ib}+\ri0$, respectively.
Here we make use of the function
\begin{equation}
\cLi2(x_1,x_2) =
\Li2\left(1-x_1 x_2\right) +\eta(x_1,x_2) \ln\left(1-x_1 x_2\right),
\label{eq:cLi}
\end{equation}
which is a specific analytical continuation of the dilogarithm $\Li2$
in two arguments $x_1$ and $x_2$, which in turn makes use of the
$\eta$~function
\begin{equation}
\eta (a,b) = 2 \pi \ri \Bigl\{ \theta (-\mathrm{Im} a)\, \theta (- \mathrm{Im} b) \,\theta
                ( \mathrm{Im} (ab))
-  \theta ( \mathrm{Im}  a) \,\theta ( \mathrm{Im} b) \,\theta (- \mathrm{Im} (ab) ) \Bigr\} .
\end{equation}
The remaining $C_0$ and $D_0$ integrals can be found in
Refs.~\cite{Dittmaier:2003bc} and \cite{Denner:2010tr}, respectively.  The
five-point integral $E_0$ can be reduced to five four-point integrals $D_0$ as,
e.g., described in Refs.~\cite{Melrose:1965kb,Denner:2002ii}.

Finally, we recall the simple, well-known substitution that translates a pure
soft IR singularity from mass regularization by the infinitesimal mass
$m_\gamma$ to regularization in $D=4-2\epsilon$ dimensions,
\begin{equation}
\ln(m_{\gamma}^2) \;\to\;
\frac{\Gamma(1+\epsilon)}{\epsilon} (4\pi\mu^2)^\epsilon
+ {\cal O}(\epsilon).
\end{equation}

\end{appendix}

\printbibliography

\end{document}